\documentclass[10pt,twocolumn,journal]{IEEEtran}

\usepackage{epsfig}
\usepackage{epstopdf}
\usepackage{citesort}
\usepackage{amsmath}
\usepackage{amssymb}
\usepackage{array}
\usepackage{color}
\usepackage{url}
\usepackage{multirow}
\usepackage{mathtools}
\usepackage{algorithm}
\usepackage{algorithmic}
\usepackage{subfigure}
\usepackage{widetext}
\usepackage{diagbox}
\usepackage{graphicx}

\newcommand{\eg}{\textit{e.g., }}
\newcommand{\ie}{\textit{i.e., }}

\newlength{\figurewidth}
\newlength{\smallfigurewidth}
\newlength{\figureheight}
\begin{document}
	\title
	{Dynamic Proximal Unrolling Network for Compressive Imaging}
	
	\author{%
		Yixiao Yang,
		Ran Tao,~\IEEEmembership{Senior Member,~IEEE},
		Kaixuan Wei,
		Ying Fu,~\IEEEmembership{Member,~IEEE}\\
		\thanks{%
			This work was supported in part by the National Natural Science Foundation of China under Grant U1833203. (Corresponding author: Ran Tao, e-mail: rantao@bit.edu.cn).}
		\thanks{%
			Y. Yang and R. Tao are with the School of Information and Electronics, Beijing Institute of Technology, Beijing 100081, China (e-mail: yixiaoyang@bit.edu.cn; rantao@bit.edu.cn). }
		\thanks{%
			K. Wei and Y. Fu are with the School of Computer Science and Technology, Beijing Institute of Technology, Beijing 100081, China (e-mail: kaixuan\_wei@bit.edu.cn; fuying@bit.edu.cn). }
	}
	
	\maketitle
	\thispagestyle{empty}
	\pagestyle{empty}
	
	\begin{abstract}
		Compressive imaging aims to recover a latent image from under-sampled measurements, suffering from a serious ill-posed inverse problem.
		Recently, deep neural networks have been applied to this problem with superior results, owing to the learned advanced image priors.
		These approaches, however, require training separate models for different imaging modalities and sampling ratios, leading to overfitting to specific settings. 
		In this paper, a dynamic proximal unrolling network (dubbed DPUNet) was proposed, which can handle a variety of measurement matrices via one single model without retraining. 
		Specifically, DPUNet can exploit both the embedded observation model via gradient descent and imposed image priors by learned dynamic proximal operators, achieving joint reconstruction.
		A key component of DPUNet is a dynamic proximal mapping module, whose parameters can be dynamically adjusted at the inference stage and make it adapt to different imaging settings.
		Experimental results demonstrate that the proposed DPUNet can effectively handle multiple compressive imaging modalities under varying sampling ratios and noise levels via only one trained model, and outperform the state-of-the-art approaches.
	\end{abstract}
	

	\section{Introduction}
	\label{sec:intro}
	
	Compressive imaging depicts a novel imaging paradigm for image acquisition and reconstruction that allows the recovery of an underlying image from far fewer measurements than the Nyquist sampling rate \cite{donoho2006compressed,liutkus2014imaging,duarte2008single,kerviche2014information}, and drives a range of practical applications, such as image or video compressive sensing \cite{liutkus2014imaging,sankaranarayanan2012cs}, compressive sensing magnetic resonance imaging (CS-MRI) \cite{lustig2007sparse,lustig2008compressed}, single-pixel imaging \cite{duarte2008single,rousset2016adaptive}, snapshot compressive imaging \cite{yuan2021snapshot}, and compressive phase retrieval (CPR) \cite{moravec2007compressive,ohlsson2012cprl}. 
	
	Mathematically, given the under-sampled measurements $y\in \mathbb{R}^{M}$ and observation model $\Phi (\cdot)$, the goal of compressive imaging is to find a solution $\hat{x}\in \mathbb{R}^{N}$, such that $y \thickapprox \Phi (\hat{x})$ and $\hat{x}$ resides the class of images. Since the sampling ratio, defined as $\frac{M}{N}$, is typically much less than one, reconstructing a unique solution from limited measurements only is difficult or impossible without proper image priors.

	\begin{figure}[!t]
		\centering
		\setlength\tabcolsep{0.8pt}
		\begin{tabular}{m{2cm}<{\centering} m{2cm}<{\centering} m{2cm}<{\centering} m{2cm}<{\centering}}
			{} & {Ground Truth} & {Initialization} & {Reconstruction}\\
			{BCS ($10\times$ compression)} &
			\includegraphics[width=\linewidth,clip,keepaspectratio]{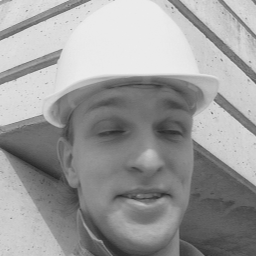} &
			\includegraphics[width=\linewidth,clip,keepaspectratio]{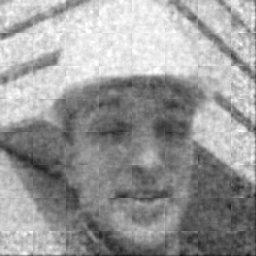} &
			\includegraphics[width=\linewidth,clip,keepaspectratio]{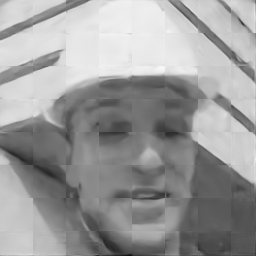} \\
			& \footnotesize  PSNR & \footnotesize 25.47 dB & \footnotesize  30.13 dB\\
			{CS-MRI ($5\times$ compression)} &
			\includegraphics[width=\linewidth,clip,keepaspectratio]{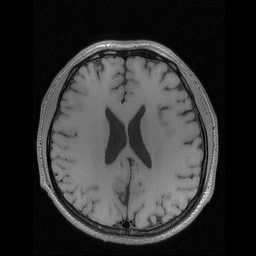} &
			\includegraphics[width=\linewidth,clip,keepaspectratio]{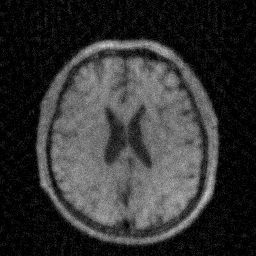} &
			\includegraphics[width=\linewidth,clip,keepaspectratio]{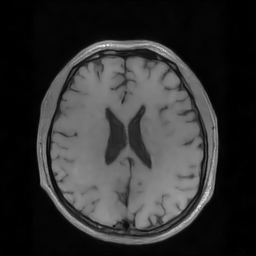} \\
			& \footnotesize  PSNR & \footnotesize 25.86 dB & \footnotesize 33.16 dB\\
			{CPR ($2\times$ compression)} &
			\includegraphics[width=\linewidth,clip,keepaspectratio]{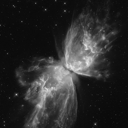} &
			\includegraphics[width=\linewidth,clip,keepaspectratio]{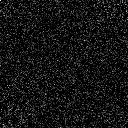} &
			\includegraphics[width=\linewidth,clip,keepaspectratio]{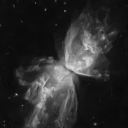} \\
			& \footnotesize  PSNR & \footnotesize 9.81 dB & \footnotesize  35.36 dB\\
		\end{tabular}
		\caption{The single trained model of our network is used to handle multiple compressive imaging modalities with different sampling ratios: block-based compressive sensing (BCS) with $10\times$ compression, CS-MRI with $5\times$ compression, and CPR with $2\times$ compression. Note that even though these inverse problems are very different, the proposed network can handle multiple imaging modalities under various imaging conditions via only one trained model without retraining.}
		\label{fig:review_results}
	\end{figure}
	
	To tackle the fundamental ill-posedness of compressive imaging, traditional methods typically exploit the measurement model knowledge and intrinsic image properties \cite{ma2008efficient,liao2008sparse,dong2014compressive,qu2014magnetic,zhang2014group,6873741,metzler2016denoising}, and solve a regularized optimization problem in an iterative scheme \cite{combettes2011proximal,afonso2010fast,beck2009fast,geman1995nonlinear,boyd2011distributed,yang2010fast}. They are effective and flexible to handle a wide variety of measurements based on the well-studied forward model and well-understood behavior but limited in unsatisfactory reconstruction quality and high computational complexity.
	
	In contrast to iterative-based methods, deep-learning-based compressive imaging approaches \cite{ongie2020deep,mousavi2015deep,iliadis2018deep,kulkarni2016reconnet}, as an alternative, employ neural networks to directly learn a mapping from measurements to latent quantities entirely based upon data. Once trained, the inference only requires a single forward of the network, without the need for time-consuming optimization and hyper-parameters selection. Nevertheless, the pure deep learning approach cannot offer the flexibility of variational methods in adapting to different imaging modalities and even sampling ratios, largely due to learning a task-specific mapping. 
	Traditionally one would require a separate deep network for each imaging setting even if there is only a tiny change, which limits its practical applications.
	
	Motivated by the observation that many iterative optimization algorithms can be truncated and unfolded into learnable deep neural networks, researchers have explored the hybrid approach (\ie deep unrolling) that combines the best of both worlds in compressive imaging \cite{gregor2010learning,borgerding2017amp,yang2016deep,zhang2018ista}. Following the unrolling, regularizers or any other free hyper-parameters such as the step size or regularization parameters, can be learned via end-to-end training, rather than being hand-crafted, meanwhile, the physical measurement model can be explicitly exploited. In this way, deep unrolling networks take the merits of interpretability and flexibility of optimization-based methods and fast inference of deep-learning-based approaches, while achieving promising reconstruction performance.
	
\begin{figure}[!t]
		\centering
		\includegraphics[width=0.9\linewidth,clip,keepaspectratio]{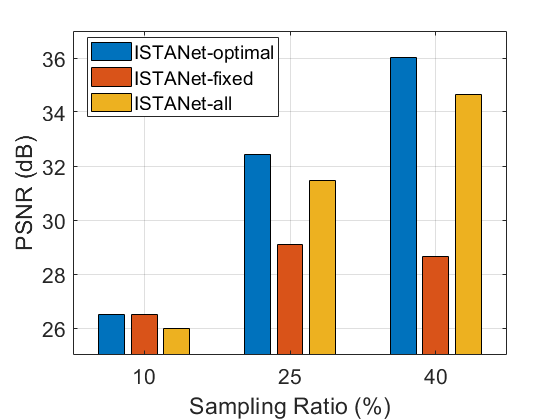}
		
		\caption{The model generalizability test of a representative deep unrolling method -- ISTANet \cite{zhang2018ista} for image compressive sensing under multiple sampling ratios. 
			``ISTANet-optimal'' denotes the (three) models trained and tested on consistent sampling ratios.  
			``ISTANet-fixed'' indicates the single model trained on the fixed sampling ratio 10\% and then tested on all sampling ratios.
			``ISTANet-all'' implies the single model trained and tested on all sampling ratios (\ie 10\%, 25\%, and 40\%).}
		
		\label{fig:generalization}
	\end{figure}

	Despite these gains, the existing deep unrolling networks still suffer from severe performance degradation when operating on a measurement matrix significantly different from the training set.
	An illustrative example of image compressive sensing is shown in Fig.~\ref{fig:generalization}, where the model trained on a fixed sampling ratio (``fixed") produces poor results when testing on unseen sampling ratios during training. While the performance can be improved when training on all sampling ratios jointly (``all"), there is still a large gap between such model and the task-specific model separately trained on each sampling ratio ("optimal"). 
	
	Basically, the ill-posedness of compressive imaging and the difficulty of learning to solve the corresponding inverse problem are very distinct under different imaging conditions. Furthermore, learned parameters of a specific unrolling network remain fixed after training, so the inference cannot adapt dynamically to other imaging parameter configurations.
	To address this issue, in this paper, a novel deep unrolling architecture was proposed, whose key part is a dynamic proximal mapping module. Specifically, this module consists of a convolutional neural network (CNN) that learns/executes proximal operators and several secondary fully connected networks that perform dynamic modification mechanisms to adjust the parameters of CNN given imaging conditions, which can be jointly trained end-to-end. In this way, fully connected networks will dynamically adjust the learned proximal operators at the inference stage and make the unrolling network adapt to different imaging settings.
	
	The effectiveness of the proposed method is verified on three representative compressive imaging applications, \ie image compressive sensing, CS-MRI, and CPR under various imaging conditions. Furthermore, the applicability of the proposed framework is explored on multiple imaging modalities with different imaging conditions. Experimental results demonstrate that the proposed method can not only effectively tackle varying imaging conditions for a specific compressive imaging task, but also be able to handle multiple imaging modalities via only \textit{one single model} simultaneously, and outperform the state-of-the-art approaches. 
	
	Our main contributions can be summarized as follows:
	\begin{itemize}
		\item We present a dynamic proximal unrolling network (dubbed DPUNet) that can adaptively handle different imaging conditions, and even various compressive imaging modalities via the only one trained model.
		\item The key part of DPUNet is to develop a dynamic proximal mapping module, which can enable the on-the-fly parameter adjustment at the inference stage and boost the generalizability of deep unrolling networks.
		\item Experimental results demonstrate DPUNet can outperform the state-of-the-art on image compressive sensing, CS-MRI, and CPR under various imaging conditions without retraining. In addition, we show the extension of DPUNet can simultaneously handle all these imaging tasks via one single trained model, with promising results.
	\end{itemize}
	
	The remainder of this paper is organized as follows. We review related compressive imaging methods in Section \ref{sec:relatedwork}. In Section \ref{sec:method}, the details of the proposed method including unrolling frameworks and dynamic proximal mapping module are presented. Section \ref{sec:Experiments} provides both experimental settings and qualitative results. In Section \ref{sec:conclusion}, we conclude the paper.
	
	\section{Related Work}
	\label{sec:relatedwork}

	
	
	Traditionally, the inverse problem of compressive imaging can be attacked by variational optimization methods by minimizing the following cost functional: 
	\begin{align}
	\mathop{\mathrm{\mathop{minimize}}}_x \quad \frac{1}{2}||y - \Phi (x) ||^2_2 +  \lambda \mathcal{R} \left( x \right),
	\label{eq:optimization}
	\end{align}
	by an iterative optimization framework, \eg the proximal gradient descent (PGD) \cite{combettes2011proximal}, the alternating direction method of multipliers (ADMM) \cite{boyd2011distributed}, the iterative shrinkage-thresholding algorithm (ISTA) \cite{beck2009fast}, and the half-quadratic splitting (HQS) \cite{geman1995nonlinear} algorithm. In the above, $R(x)$ indicates the regularization term associated with the prior knowledge of images, which alleviates the ill-posedness of compressive imaging. 
	Many image priors have been designed in the imaging community over the past decades. The well-known examples include structured sparsity \cite{mun2009block,liao2008sparse,ravishankar2010mr,9361634}, group sparse representation (GSR) \cite{zhang2014group} and nonlocal low-rank \cite{dong2014compressive,mairal2009non,qu2014magnetic}. 
	
	In contrast to handcrafted priors, deep-learning-based compressive imaging approaches have been proposed and demonstrated promising reconstruction results with fast inference speed \cite{mousavi2015deep,mousavi2017learning,iliadis2018deep,kulkarni2016reconnet,adler2016deep,9484349,9470780,shi2019scalable}. 
	A stacked denoising auto-encode (SDA) was first applied to learn statistical dependencies from data, improving signal recovery performance \cite{mousavi2015deep}. Fully connected neural networks were proposed for image and video block-based compressive sensing (BCS) reconstruction \cite{iliadis2018deep,adler2016deep}. Further several CNN-based approaches were developed, which learn the inverse map from compressively sensed measurements to reconstructed images \cite{mousavi2017learning,kulkarni2016reconnet,shi2019scalable}.
	The basic idea of these works is designing a neural network $f_{NN}(y,\Theta)$ to directly perform the inverse mapping from observed measurements $y$ to desired images $x$ by learning network parameters $\Theta$ based upon the training dataset $\Gamma$:
	\begin{align}
	\hat{\Theta}=\mathop{\mathrm{\mathop{argmin}}}_\Theta \quad \frac{1}{|\Gamma|}\sum_{(y,x)\in \Gamma} \mathcal{L}(x,f_{NN}(y,\Theta)),
	\label{eq:cnn}
	\end{align}
	where $\mathcal{L}$ is a loss function. However, these network architectures are predefined and fixed for specific problems, which usually cannot adapt to others without retraining. In addition, designing neural networks could be often considered as much an art as a science, without clear theoretical guidelines and domain knowledge \cite{hershey2014deep}. 
	
	Inspired by the interpretability and flexibility of traditional approaches, an emerging technique called deep unrolling or unfolding has been applied to compressive imaging \cite{gregor2010learning,borgerding2017amp,yang2016deep,zhang2018ista}.
	By unrolling the iterative optimization framework and introducing learnable parameters $\Theta$, deep unrolling networks can be learned by minimizing the following empirical risk:
	\begin{align}
	\hat{\Theta}=\mathop{\mathrm{\mathop{argmin}}}_\Theta \quad \frac{1}{|\Gamma|}\sum_{(y,x)\in \Gamma} \mathcal{L}(x,f_{NN}(y, \Phi, \Theta)),
	\label{eq:unrolling}
	\end{align}
	where $\Phi$ is the given physical measurement model.
	Preliminary attempts focused on learning fast approximation of specialized iterative solvers attached to well-designed priors. The computation schemes of the resulting networks are concise with the original solvers, but with fixed numbers of iterations and some untied/unshared parameters across layers/iterations. 
	Well-known examples include learned shrinkage-thresholding algorithms (LISTA) \cite{gregor2010learning} and learned approximate message passing (LAMP) algorithms \cite{borgerding2017amp}.  
	A similar idea has also been applied to the alternating direction method of multipliers (ADMM) solver for CS-MRI \cite{sun2016deep}, but with the goal to design powerful networks (ADMM-Net) rather than approximating variational methods. In contrast with encoding the sparsity on linear transform domain to network \cite{sun2016deep}, ISTA-Net \cite{zhang2018ista} goes beyond that to nonlinear transform domain sparsity. Moreover, these learned sparsity constraints can be further released into purely data-driven prior to boost the performance  \cite{NIPS2017_6774,aggarwal2018modl,dong2018denoising,wang2019deep,zhang2020deep}. Nevertheless, few works consider the performance degradation of deep unrolling networks under mismatched imaging settings during inference, owing to the constant network parameters. To the best of our knowledge, this paper is the first work that presents a novel dynamic deep unrolling network to solve this issue.
	
	Another approach that can combine the benefits of both deep learning and optimization methods
	is called plug-and-play (PnP) priors \cite{venkatakrishnan2013plug}. Its core idea is to plug a denoiser into the iterative optimization such that the image prior is implicitly defined by the denoiser itself. 
	Many deep-learning-based denoisers have been utilized in the PnP framework to resolve compressive imaging problems, without the need for task-specific training \cite{Chang_2017_ICCV,metzler2018prdeep,Meinhardt_2017_ICCV,wei2020tuning-free,zhang2017learning,metzler2016bm3d-prgamp:}. 
	The main strength of PnP over deep unrolling is the generalizability (only one network is required to handle various compressive imaging problems \cite{Chang_2017_ICCV}), but
	its performance often lags behind the deep unrolling network due to the lack of end-to-end training \cite{zhang2020deep}. Meanwhile, it also faces the limitations of high computational complexity and difficulties of tuning parameters \cite{wei2020tuning-free,wei2020tfpnp}.
	In this work, we propose a dynamic proximal unrolling approach that improves the generalizability of deep unrolling considerably, while still enjoying the joint optimization of parameters via end-to-end learning.

	\begin{figure*}[htb]
		\centering
		\includegraphics[width=.9\linewidth]{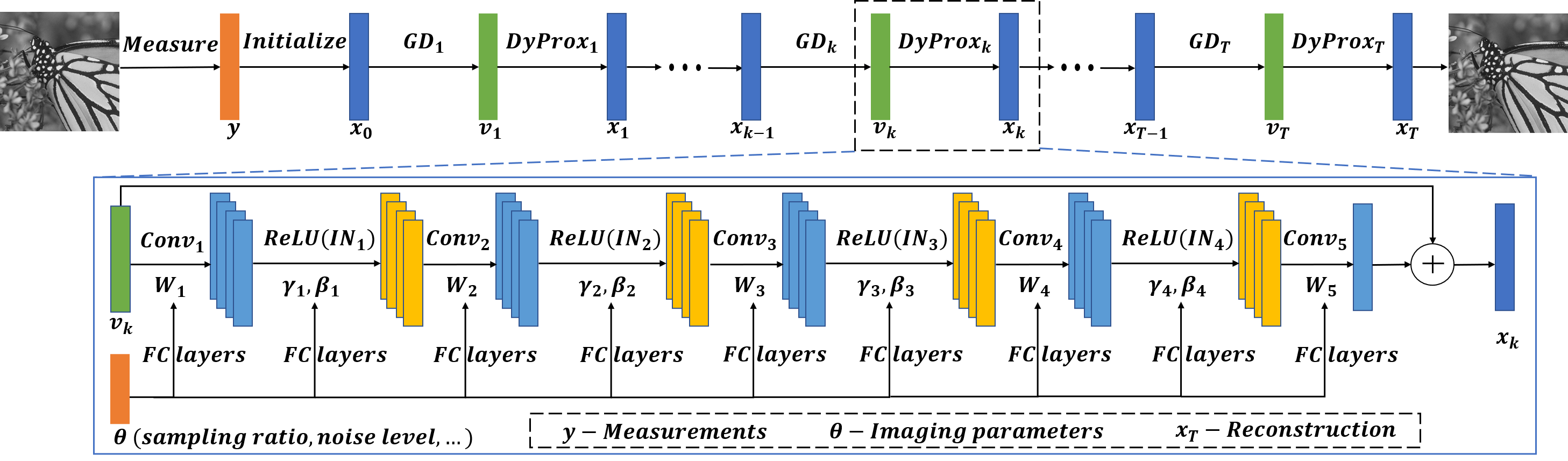}
		\caption{Overview of the proposed dynamic proximal unrolling network. Specifically, our network is unrolled by $T$ iterations (empirically set to 10), and each iteration includes the gradient descent ($GD$) and the proposed dynamic proximal ($DyProx$) mapping module, which correspond to Eq.~\eqref{eq:gd} and Eq.~\eqref{eq:prox}. The $DyProx$ consists of a deep convolutional neural network (CNN) and several fully connected (FC) layers. The CNN is designed as a combination of five convolutions (Conv) layers, each of which is separated by an instance normalization (IN) layer and a ReLU non-linearity. An identity skip connection is added between the input and output. The inputs of FC layers are imaging parameters $\theta$ such as sampling ratio and noise level, and the outputs are weights $W$ of convolution (Conv) and affine parameters $\gamma, \beta$ of instance normalization (IN). At the inference stage, given the imaging parameters, FC layers will adaptively determine parameters of CNN, which contribute to reconstructions together with gradient descent steps from given measurements.}
		\label{fig:framework}
	\end{figure*}

	\section{Method}
	\label{sec:method}
	
	In this section, we first introduce the iterative proximal optimization algorithm and the corresponding deep unrolling network. Then, we describe the dynamic proximal mapping module -- a core component of the proposed dynamic proximal unrolling network. What's more, we present the applicability of the proposed module to other optimization frameworks. An overview of the proposed network is shown in Fig.~\ref{fig:framework}. 
	
	\subsection{Proximal optimization and deep unrolling}
	
	In this paper, we adopt the proximal gradient descent (PGD) algorithm~\cite{combettes2011proximal} for solving Eq.\eqref{eq:optimization}, which provides a general and efficient scheme to split the data-fidelity term and regularization term by alternating between the gradient descent step and proximal mapping step. Starting from the initial point $x_0$, the whole iterations' process of PGD can be written as
	\begin{align} 
	&z_{k} = x_{k-1} - r_{k-1}\nabla D\left(x_{k-1}\right) \label{eq:gd},\\
	&x_{k} = \mathrm{Prox}_{\lambda \mathcal{R}} \left(z_{k} \right)\label{eq:prox},
	\end{align}
	where $k$ denotes the iteration index, $D(x) = \frac{1}{2}||y - \Phi (x) ||^2_2 $ is the data-fidelity term, $\nabla$ denotes the vector differential operator that calculates the gradient of a given function, and $r_{k}$ is the step-size of the $k$-th iteration. Eq.\eqref{eq:gd} can be understood as one-step gradient descent for dealing with the data-fidelity term. And $\mathrm{Prox}_{\lambda \mathcal{R}} (z):= \mathop{\mathrm{argmin}}_x \big\{ \mathcal{R}(x) + \tfrac{1}{2\lambda} \| x - z \|_2^2 \big\}$ denotes the proximal operator for handling the regularization term \cite{afonso2010fast,beck2009fast} in Eq.\eqref{eq:prox}. 
	
	
	
	Given the modular nature of the proximal optimization framework, the overall iterative procedure can be truncated and unrolled into a trainable reconstruction network by replacing all instances of the proximal operator $\mathrm{Prox}_{\lambda \mathcal{R}} (\cdot)$ with a trainable deep convolutional neural network (CNN). The CNN is expected to learn/perform the proximal operator in Eq.\eqref{eq:prox}, which can project the corrupted image into the clean image manifold.
	In this paper, we design a simple yet effective CNN consisting of five convolution (Conv) layers separated by an instance normalization (IN) layer~\cite{ulyanov2017improved} and a rectified linear unit (ReLU). IN was first proposed in style transfer and has shown significant improvement by normalizing feature statistics~\cite{ulyanov2017improved}. To stabilize training~\cite{he2016identity}, an identity skip connection is built between the input and output of the CNN. Mathematically, in the $k$-th iteration, the CNN maps ${z}_{k}$ to $x_{k}$ through
	$$f_{0} = {z}_{k},$$
	$$f_{j} = \mathrm{ReLU}(\mathrm{IN}_{j}(W_{j}*f_{j-1})), \,\,\, j=1,2,3,4,$$
	$$x_{k} = W_{5}*f_{4} + f_{0},$$
	where $W_{j}$ represents the filters of convolution with a certain kernel and stride size, $*$ denotes the convolution operation, and the $\mathrm{IN}_{j}$ represents the $j$-th instance normalization operation as
	$$\mathrm{IN}_{j}(x) = \gamma_{j}(\frac{x-u(x)}{\sigma(x)}+\beta_{j}),$$
	where $\gamma_{j}$ and $\beta_{j}$ are affine parameters learned from data; $u(x)$ and $\sigma(x)$ are the mean and standard deviation of the input $x$, computed across spatial dimensions independently for each feature channel and each sample.
	
	
		
	\subsection{Dynamic proximal mapping module}
	
\begin{figure}[t]
		\centering
		\includegraphics[width=.9\linewidth]{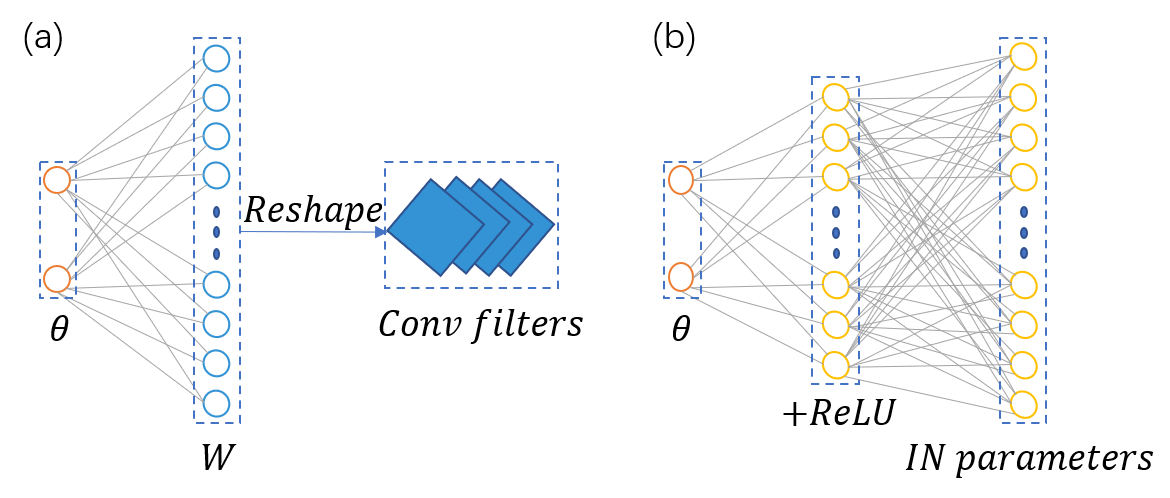}
		\caption{Illustration of the fully connected networks. Subfigure (a) illustrates the fully connected layer for convolution layers. The imaging parameter $\theta$ is an auxiliary input that feeds into a single fully-connected layer that outputs the weights $W$ of convolution filters. Subfigure (b) describes two fully connected layers separated by a ReLU for instance normalization (IN) layers. The output is affine parameters of IN.}
		\label{fig:mlp_block}
	\end{figure}

	Given the training data, the resulting network can be trained end-to-end to learn the proximal mapping. However, the trainable parameters (\ie $W_{j}, \gamma_{j}, \beta_{j}$) of the CNN are usually fixed once trained, while different imaging conditions such as varying sampling-ratios or noise-levels all affect the performance of the learned proximal operator. 
	To overcome this limitation, we make a step forward by proposing a dynamic proximal mapping module that can dynamically adjust the parameters of CNN according to different imaging conditions, to perform adaptively proximal operators. To this end, we use a set of secondary fully connected networks, whose inputs are imaging parameters $\theta$ such as the sampling ratio and noise level, and outputs are the parameters ($W_{j}, \gamma_{j}, \beta_{j}$) of CNN, shown in Fig.~\ref{fig:mlp_block}.
	The fully connected networks aim to generate and update the parameters of CNN, \ie the weights of convolution filters and affine parameters of IN layers. During the training stage, the proximal CNN and fully connected networks can be jointly trained. In the inference stage, given different imaging parameters, fully connected networks will adaptively adjust the learned proximal operator represented by the CNN, thus improving the representation capability.
	
	Specifically, the imaging parameter is an auxiliary input that feeds into fully connected layers with the final layer outputting the convolution filter weights. The output then is reshaped into a 4D tensor of convolution filter weights and convolved with the input image. Based on empirical results in Section \ref{sec:Experiments}, we employ a single fully-connected layer to directly learn the weights of convolution filters, which can be written as
	\begin{equation}
	W_{j} = A_{j} \theta + b_{j}, j=1,...,5,
	\label{eq:dynamic weight}
	\end{equation}
	where $\theta\in \mathbb{R}^l$ is the $l$ imaging parameters related to the imaging settings, $W_{j}\in \mathbb{R}^{m_j}$ represents the weights ($m_j$ denotes the total number of parameters) of the $j$-th convolution layer in CNN, and $A_{j}\in \mathbb{R}^{m_j\times l}, b_{j}\in \mathbb{R}^{m_j}$ are the weight and bias of the corresponding fully-connected layer.
	
	\textit{In essence, these fully connected networks introduce a dynamic modulation mechanism for the weights of convolution with imaging parameters.} Considering a special case, it is mathematically equivalent to directly learn the weights of convolution without fully-connected layers when the imaging parameters $\theta = 0$, \ie $W_{j} = b_{j}, j=1,...,5$. When the imaging parameters are non-zero, the fully connected networks will set up a connection between the CNN weight space and imaging parameters. 
	
	For instance normalization layers, we employ two fully-connected layers separated by a ReLU (non-linearity) for the affine parameters of IN, respectively, which can be written as
	\begin{equation}
	Q_{j} = A_{j2} \mathrm{ReLU}(A_{j1} \theta + b_{j1}) + b_{j2}, j=1,...,4,
	\end{equation}
	where $Q_{j}\in \mathbb{R}^m$ represents the parameters $\gamma_{j}\in \mathbb{R}^m$ or $\beta_{j}\in \mathbb{R}^m$ of the $j$-th instance normalization, $A_{j1}\in \mathbb{R}^{m\times l}, A_{j2}\in \mathbb{R}^{m\times m}, b_{j1}\in \mathbb{R}^m, b_{j2}\in \mathbb{R}^m$ are the weight and bias of the corresponding two fully-connected layers. Note that continually increasing the fully-connected layers can still improve the performance, but with more learnable parameters and computational complexity (See Section \ref{sec:Experiments}).
	
	\subsection{Dynamic proximal unrolling network}
	
	In essence, the structure of the proposed dynamic proximal unrolling network (dubbed DPUNet) is derived from the truncated proximal gradient descent algorithm, combined with a dynamic proximal mapping module.
	
	When given training data, the inputs of the proposed DPUNet are the measurements $y$, corresponding physical forward model $\Phi$ and imaging parameters $\theta$, which are sent to the reconstruction process and fully connected networks, respectively. So the proposed DPUNet can be trained by minimizing the following empirical risk:
	\begin{align}
	\hat{\Theta}=\mathop{\mathrm{\mathop{argmin}}}_\Theta \quad \frac{1}{|\Gamma|}\sum_{(y,x)\in \Gamma} \mathcal{L}(x,f_{NN}(y, \Phi, g(\theta; \Theta))),
	\label{eq:dpun}
	\end{align}
	where $g(\cdot; \Theta)$ is the operation for generating the dynamic parameters via fully connected layers, and $\mathcal{L}(\cdot,\cdot)$ is exploited by the pixel-wise $L_2$ loss.
	
	Given imaging parameters, fully connected networks will update the parameters of CNN to execute the proximal operator at each iteration. Finally, the outputs of the proposed network are reconstructed images, which are used to compute loss with the ground-truth. By back-propagation, the trainable parameters of our network, including the weight and bias of fully connected networks and the gradient step sizes, can be jointly optimized.
	
	
	
	Once the model is trained, given the imaging parameters and measurements, the fully connected networks can adaptively generate parameters of CNN and the unrolling network performs the reconstructions. In this way, we can dynamically adjust the learned proximal operator based on the designed fully connected networks at the inference stage, thus enabling the handling of different imaging settings and continuous parameter control.
	
	\subsection{Other optimization frameworks}
	
	A large variety of first-order proximal algorithms have been developed for solving Eq.\eqref{eq:optimization} efficiently \cite{combettes2011proximal,boyd2011distributed,beck2009fast,geman1995nonlinear}. In this paper, we present other two representative proximal optimization frameworks (\ie HQS and ADMM) to construct dynamic proximal unrolling networks. It has been observed that both of them perform well with the proposed dynamic proximal mapping module. 
	
	\medskip
	\noindent {\bf Unrolled-HQS.~} 
	HQS tackles Eq.\eqref{eq:optimization} by introducing an auxiliary variable $z$, leading to iteratively solving subproblems for $z$ and $x$ as
	\begin{align} 
	z_{k} &= \mathop{\mathrm{argmin}}_z \big\{ \frac{1}{2}||y - \Phi (z) ||^2_2 + \tfrac{\mu_k}{2} \| z - x_{k-1} \|_2^2 \big\} \label{eq:hqs-z},\\
	x_{k} &= \mathop{\mathrm{argmin}}_x \big\{\lambda \mathcal{R}(x) + \tfrac{\mu_k}{2} \| z_k - x \|_2^2 \big\}, \label{eq:hqs-x}
	\end{align}
	where $k$ denotes the iteration index, $\mu_k$ indicates the penalty parameter. 
	
	Assuming that $\Phi(\cdot)$ is a linear measurement model, we still adopt the one-step gradient descent and proximal operator to deal with Eq.\eqref{eq:hqs-z} and Eq.\eqref{eq:hqs-x}, respectively, which can be written as
	\begin{align} 
	&z_{k} = z_{k-1} - r_{k-1}(\Phi^{T}(\Phi z_{k-1}-y)+\mu_k (z_{k-1}-x_{k-1})) \label{eq:hqs-1},\\
	&x_{k} = \mathrm{Prox}_{\frac{\lambda}{\mu_k} \mathcal{R}} \left(z_{k} \right) \label{eq:hqs-2},
	\end{align}
	where  $\Phi^{T}$ denotes the transpose of the sampling matrix, $\mathrm{Prox}$ denotes the proximal operator, and $r_{k}$ is step-size at $k$-th iteration.
	
	Once the proximal optimization is determined, the next step is to unroll the iterative process into a dynamic proximal unrolling network by introducing the designed dynamic proximal mapping module. In this way, the free hyper-parameters (\ie $r_k,\mu_k$) of HQS and dynamic proximal mapping module can be jointly learned via end-to-end training.
	
	\medskip
	\noindent {\bf Unrolled-ADMM.~} 
	Eq.\eqref{eq:optimization} can also be solved by ADMM, whose iterations can be written as
	\begin{align} 
	&x_{k} = \mathrm{Prox}_{\lambda \mathcal{R}} \left(z_{k-1} - u_{k-1} \right) \label{eq:admm-1},\\
	&z_{k} = \mathrm{Prox}_{\frac{1}{\mu_k} \mathcal{D}} \left(x_{k} + u_{k-1} \right)\label{eq:admm-2}, \\
	&u_{k} = u_{k-1} + x_{k} - z_{k}\label{eq:pnp-admm-3},
	\end{align}
	where $\mathcal{D}(x) = \frac{1}{2}||y - \Phi (x) ||^2_2 $, $\mu_k$ still indicates the penalty parameter.
	
	Similarly, we use the one-step gradient descent to tackle Eq.\eqref{eq:admm-2}. Unrolled-ADMM can be derived by unrolling the corresponding iterative optimization and replacing the proximal operator with the designed dynamic proximal mapping module.
	
	Based on empirical results, we demonstrate the proposed dynamic proximal mapping module can be embedded into different proximal optimization frameworks, and boost the generalizability of deep unrolling network. In this paper, we adopt the proximal gradient descent framework as the final choice, owing to its conciseness and effectiveness. More experimental details are presented in Section \ref{sec:Experiments}.
	
	\section{Experiments}
	\label{sec:Experiments}
	
	In this section, we mainly focus on three representative compressive imaging modalities: image compressive sensing, compressive sensing magnetic resonance imaging (CS-MRI), and compressive phase retrieval (CPR), and detail experiments to evaluate the proposed method. We first describe the experimental setting including both physical measurement models and implementation details. Then we compare our method against prior art on different tasks under varying imaging conditions and provide an in-depth discussion of the proposed method. Finally, we present an extension of DPUNet to simultaneously handle all these compressive imaging modalities via one single model.
	
	\subsection{Experimental Setting}
	
	\subsubsection{Image CS}
	
	Image compressive sensing (CS) is a popular and well-studied linear inverse problem, which enables image or video capturing under a sub-Nyquist sampling rate~\cite{liutkus2014imaging,sankaranarayanan2015video}. Following common practices in previous CS work, we focus on block-based compressive sensing (BCS) tasks to validate the proposed method. Given the sampling ratio, the measurement matrix $\Phi$ is a random Gaussian matrix and the measurement is generated by $y=\Phi x$, where $x$ is the vectorized version of an image block with a size of $33\times 33$.
	For BCS, we adopt the data-fidelity term $D(x) = \frac{1}{2}||y - \Phi x ||^2_2 $, and compute its gradient $\nabla D(x) = \Phi^T (\Phi x-y)$. Here $\Phi^T$ denotes the transpose of $\Phi$. For initialization, we adopt the linear mapping method, same as ISTA-Net~\cite{zhang2018ista}. 
	For a fair comparison, we utilize the same training data pairs in ISTA-Net~\cite{zhang2018ista} for training and widely used public dataset Set11 \cite{Set12} for testing. 
	In order to handle various imaging conditions, we simulate to generate the BCS measurement of each patch with the sampling ratio $\eta$ uniformly sampled from \{0.01,0.04,0.1,0.25,0.4,0.5\} and add Gaussian noise with the noise level $\alpha$ uniformly sampled from [0,50], rather than a specific setting. Then we take $\{\eta,\alpha\}$ as imaging parameters during training because there will always be noise in real CS measurements.
	
	\begin{table*}[!t]
		\centering
		\small
		\caption{Average PSNR (dB) performance comparisons for BCS reconstructions on Set11 under different imaging conditions ($\eta$ and $\alpha$ denote the sampling ratio and noise level respectively). The best results are labeled in \textbf{bold} and the second are \underline{underlined}.}
		\label{tab:csbenchmark}
		
		\begin{tabular}{|c|c|ccccccc|}
			\hline  
			\centering
			{\textsc{Sampling Ratio}} & {\textsc{Noise Level}} & {TVAL3} & {D-AMP} & {NLR-CS} & {ReconNet} & {ISTA-Net} & {ISTA-Net$^+$} & {DPUNet}\\ 
			\hline
			\centering
			\multirow{3}{*}{$\eta=50\%$} & {$\alpha=10$} & {26.83} & {30.63} & {26.12} & {25.60} & {30.74} & \underline{31.42} & {\textbf{31.99}}\\
			& {$\alpha=30$} & {19.08} & {25.92} & {19.84} & {24.39} & {26.36} & \underline{26.87} & {\textbf{27.20}}\\
			& {$\alpha=50$} & {14.88} & {23.56} & {15.62} & {22.34} & {23.85} & \underline{24.30} & {\textbf{24.79}}\\
			\hline
			\centering
			\multirow{3}{*}{$\eta=40\%$} & {$\alpha=10$} & {26.31} & {29.29} & {25.16} & {25.19} & {28.73} & \underline{30.53} & {\textbf{31.20}}\\
			& {$\alpha=30$} & {19.33} & {25.08} & {20.05} & {23.78} & {25.42} & \underline{26.21} & {\textbf{26.59}}\\
			& {$\alpha=50$} & {15.38} & {22.84} & {16.08} & {21.56} & {23.14} & \underline{23.78} & {\textbf{24.17}}\\
			\hline
			\centering
			\multirow{3}{*}{$\eta=30\%$} & {$\alpha=10$} & {25.57} & {27.58} & {24.42} & {24.23} & {27.07} & \underline{29.45} & {\textbf{30.04}}\\
			& {$\alpha=30$} & {19.47} & {23.93} & {20.51} & {23.12} & {24.46} & \underline{25.28} & {\textbf{25.70}}\\
			& {$\alpha=50$} & {15.81} & {22.00} & {16.70} & {21.24} & {22.30} & \underline{22.89} & {\textbf{23.29}}\\
			\hline
			\centering
			\multirow{3}{*}{$\eta=10\%$} & {$\alpha=10$} & {21.91} & {20.87} & {21.48} & {21.91} & {23.33} & \underline{24.63} & {\textbf{25.40}}\\
			& {$\alpha=30$} & {18.83} & {19.68} & {20.23} & {20.92} & {21.30} & \underline{21.85} & {\textbf{22.26}}\\
			& {$\alpha=50$} & {16.34} & {18.53} & {17.83} & {19.19} & {19.47} & \underline{20.02} & {\textbf{20.32}}\\
			\hline
			\centering
			\multirow{3}{*}{$\eta=1\%$} & {$\alpha=10$} & {11.26} & {5.20} & {11.43} & {17.04} & {17.17} & \underline{17.25} & {\textbf{17.27}}\\
			& {$\alpha=30$} & {10.90} & {5.19} & {10.93} & {16.44} & \underline{16.47} & {16.46} & {\textbf{16.58}}\\
			& {$\alpha=50$} & {10.55} & {5.18} & {10.51} & {15.45} & \underline{15.45} & {15.44} & {\textbf{15.79}}\\
			\hline
		\end{tabular}
		
	\end{table*}

	\begin{figure*}[t]
		\centering
		\setlength\tabcolsep{0.8pt}
		\begin{tabular}{cccccccc}
			{TVAL3} & {D-AMP} & {NLR-CS} & {ReconNet} & {ISTA-Net} & {ISTA-Net$^+$} & {DPUNet} & {GroundTruth}\\
			\includegraphics[width=.12\linewidth,clip,keepaspectratio]{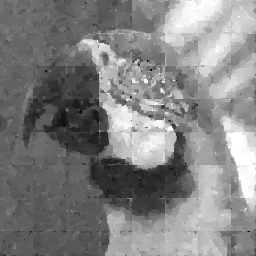} &
			\includegraphics[width=.12\linewidth,clip,keepaspectratio]{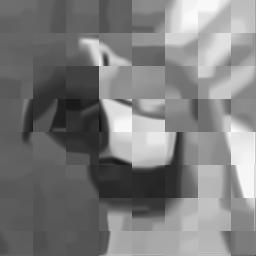} &
			\includegraphics[width=.12\linewidth,clip,keepaspectratio]{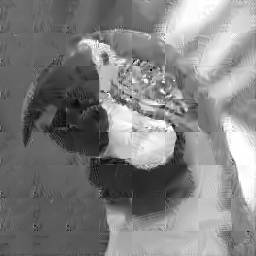} &
			\includegraphics[width=.12\linewidth,clip,keepaspectratio]{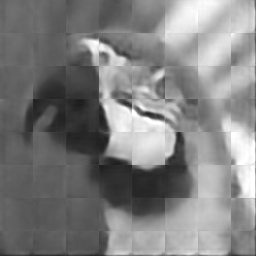} &
			\includegraphics[width=.12\linewidth,clip,keepaspectratio]{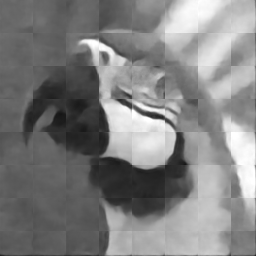} &
			\includegraphics[width=.12\linewidth,clip,keepaspectratio]{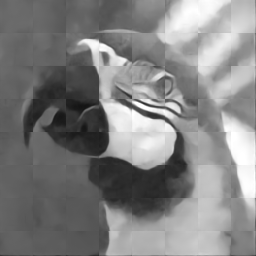} &
			\includegraphics[width=.12\linewidth,clip,keepaspectratio]{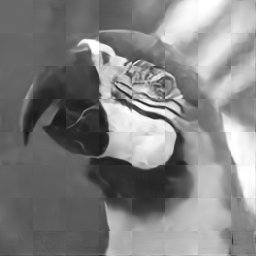} &
			\includegraphics[width=.12\linewidth,clip,keepaspectratio]{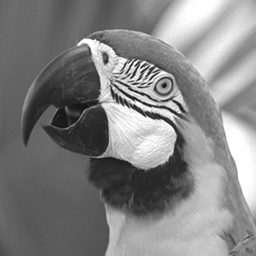} \\
			\footnotesize  22.24 & \footnotesize 21.39 & \footnotesize  23.37 & \footnotesize  22.67 & \footnotesize  23.73 & \footnotesize  24.86 & \footnotesize  25.75 & \footnotesize  PSNR\\
			\includegraphics[width=.12\linewidth,clip,keepaspectratio]{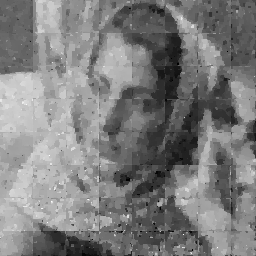} &
			\includegraphics[width=.12\linewidth,clip,keepaspectratio]{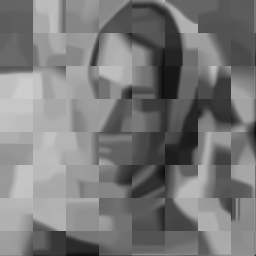} &
			\includegraphics[width=.12\linewidth,clip,keepaspectratio]{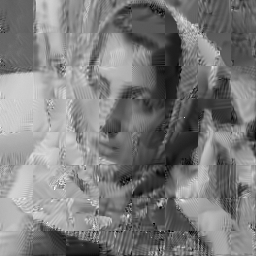} &
			\includegraphics[width=.12\linewidth,clip,keepaspectratio]{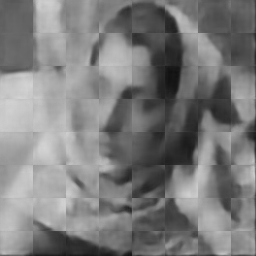} &
			\includegraphics[width=.12\linewidth,clip,keepaspectratio]{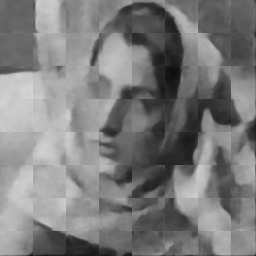} &
			\includegraphics[width=.12\linewidth,clip,keepaspectratio]{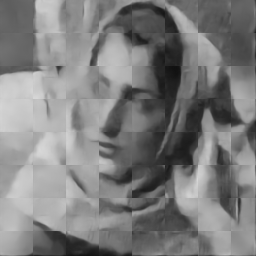} &
			\includegraphics[width=.12\linewidth,clip,keepaspectratio]{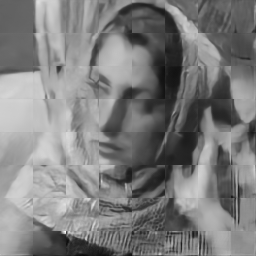} &
			\includegraphics[width=.12\linewidth,clip,keepaspectratio]{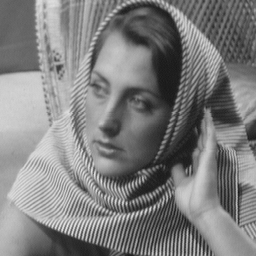} \\
			\footnotesize  21.38 & \footnotesize 20.96 & \footnotesize  22.58 & \footnotesize  21.79 & \footnotesize  22.27 & \footnotesize  22.66 & \footnotesize  23.20 & \footnotesize  PSNR\\
		\end{tabular}
		\caption{Reconstructed images and corresponding PSNRs (dB) from BCS measurements under imaging conditions ($\eta=10\%$, $\alpha=10\%$) with seven image CS algorithms.}
		\label{fig:csbenchmark}
	\end{figure*}
	
	\subsubsection{CS-MRI}
	
	CS-MRI is an advanced technique for fast MRI through reconstruction of MR images from much fewer under-sampled measurements in k-space (\ie Fourier domain). Following the common practices, we adopt the measurement matrix with the form $\Phi = PF$, where $F$ represents the 2-dimensional Fourier transform and $P$ is an under-sampling matrix taken as a commonly used pseudo radial sampling mask. We use the same training and testing brain medical images as ISTANet \cite{zhang2018ista}. For initialization, we project the under-sampled Fourier measurements to the image domain via the inverse Fourier transform. For each image, we simulate to generate subsampled measurements in Fourier domain with the sampling ratio $\eta$ uniformly sampled from \{0.2,0.3,0.4,0.5\} and add Gaussian noise $\epsilon$ with the noise level $\alpha$ uniformly sampled from [0,50].
	
	\subsubsection{CPR}
	
	Compressive phase retrieval (CPR) is a representative non-linear inverse problem, concerned with the recovery of an underlying image from only the subsampled intensity of its complex transform.  Mathematically, the measurements of CPR can be written as $y = \left| Ax \right|^2 + \epsilon$ with $\epsilon \sim \mathcal{N} (0, \alpha^2 Diag(\left|Ax\right|^2))$, where the term $\alpha$ controls the noise level in this problem \cite{yeh2015experimental,metzler2018prdeep}. The data-fidelity term adopts the amplitude loss function $D(x) = \frac{1}{2}||\sqrt{y}-\left|Ax\right| ||^2_2 $ in \cite{yeh2015experimental}. Notice $D(x)$ involves complex number operations, we therefore adopt the Wirtinger derivatives \cite{candes2015phase} to compute its gradient, \ie $\nabla D(x) = A^H \left( (|Ax|-\sqrt{y})\circ\frac{Ax}{|Ax|} \right)$, where $\circ$ denotes the Hadamard (element-wise) product and $A^H$ denotes the conjugate transpose of $A$.
	We test CPR methods under simulated coded Fourier measurements, where the measurement matrix with the form $A = J\mathcal{F}M$, where $F$ represents the 2-dimensional Fourier transform, $M$ is diagonal matrices with nonzero elements drawn uniformly from a unit circle in the complex plane \cite{metzler2018prdeep} and $J$ is an $m\times n$ matrix made from randomly sampled rows of an $n\times n$ identity matrix. Here we initialize $x_0$ with a vector of ones that worked sufficiently well.
	To train the network, we follow the common practice that uses 160000 overlapping patches (with size $64 \times 64$) cropped from 400 images from the BSD dataset \cite{martin2001a}. For each patch, we simulate to generate coded Fourier measurements with the sampling ratio $\eta$ uniformly sampled from \{0.3,0.4,0.5\} and add Poisson shot noise $\epsilon$ with the noise level $\alpha$ uniformly sampled from [0,50]. 
	
	The inputs of our network are the measurements and corresponding imaging parameters. To keep consistent magnitude of back-propagated gradient, we normalize the maximum value of imaging parameters to one. We train all models using pixel-wise $L_2$ loss and Adam optimizer with PyTorch on one Nvidia GeForce GTX 2080 Ti GPU. The models of BCS are trained in 200 epochs with batch size 64 and learning rate $10^{-4}$. The model of CS-MRI is trained in 200 epochs with batch size 4 and learning rate $10^{-4}$. The model of CPR is trained in 20 epochs with batch size 20 and learning rate $10^{-3}$.
	
    \subsection{Performance Comparison}
       
    \subsubsection{Validations on BCS}
    
    To verify the performance of the proposed method, we mainly compare it against three classic CS approaches, namely TVAL3 \cite{li2013efficient}, D-AMP \cite{metzler2016denoising}, NLR-CS \cite{dong2014compressive}, the learning-based approach ReconNet \cite{kulkarni2016reconnet} and the representative deep unrolling approach ISTA-Net \cite{zhang2018ista}. Recent works (\eg SCSNet \cite{shi2019scalable} and OPINE-Net \cite{zhang2020optimization}) use data-driven learned sampling matrix to achieve very high performance~\footnote{Despite these gains induced by the data-driven sampling matrix, the existing SOTA methods still suffer from the lack of generalization and universality. For example, OPINE-Net is designed for each sampling ratio owing to predefined CNN filters of Sampling Subnet and needs to be separately trained. SCSNet also requires to update the greedy searching strategy for different sampling ratios.}. For a fair comparison, in this paper, we only consider the competing methods using a fixed sampling matrix. We use the code made available by the respective authors on their websites. For end-to-end training methods, such as ReconNet and ISTA-Net, we train the corresponding model for each sampling ratio setting. Table~\ref{tab:csbenchmark} shows all methods' performance on the Set11 public dataset under various sampling ratios and noise levels. It can be observed that our method outperforms other methods under various sampling ratios via one single trained model. By contrast with ISTA-Net and ReconNet, our method avoids the cumbersome retraining requirement, which is beneficial for real applications. In addition, it is worth mentioning that our method can generalize well to the unseen case during training setting, as $\eta = 30\%$. We show the reconstructions of all algorithms under imaging conditions $\eta = 10\%, \alpha = 10$, in Fig.~\ref{fig:csbenchmark}. It can be seen that DPUNet produces more accurate and clearer reconstructed images than other competing algorithms.
          
    
    \begin{table*}[htbp]
    	\centering
    	\caption{Average PSNR (dB) and Run Time (s) performance comparisons for CS-MRI on 50 brain medical images under different imaging conditions ($\eta$ and $\alpha$ denote the sampling ratio and noise level respectively). The best results are labeled in \textbf{bold} and the second-best results are \underline{underlined}.}
    	\label{tab:csmribenchmark}
    	
    	\begin{tabular}{|c|ccc|ccc|ccc|ccc|c|}
    		\hline  
    		\centering 
    		\multirow{2}{*}{\textsc{Algorithm}} & \multicolumn{3}{c|}{\textsc{$\eta=20\%$}} & \multicolumn{3}{c|}{\textsc{$\eta=30\%$}} & \multicolumn{3}{c|}{\textsc{$\eta=40\%$}} & \multicolumn{3}{c|}{\textsc{$\eta=50\%$}} & {\textsc{Time}}\\ \cline{2-13}
    		~ & {\textsc{$\alpha=10$}} & {30} & {50} & {$\alpha=10$} & {30} & {50} & {$\alpha=10$} & {30} & {50} & {$\alpha=10$} & {30} & {50} & {\textsc{CPU/GPU}}\\  
    		\hline  
    		\centering   
    		{RecPF} & {31.41} & {25.06} & {21.33} & {32.47} & {24.76} & {20.61} & {33.05} & {24.37} & {19.96} & {33.46} & {24.00} & {19.41} & {0.39s/-}\\ 
    		{FCSA} & {30.71} & {24.11} & {19.19} & {31.31} & {23.07} & {17.93} & {31.28} & {22.28} & {17.21} & {31.29} & {21.78} & {16.77} & {0.64s/-}\\ 
    		{DMRI} & {25.77} & {20.23} & {22.06} & {32.15} & {24.66} & {20.69} & {31.80} & {23.71} & {19.64} & {31.35} & {22.92} & {18.80} & {10.28s/-}\\ 
    		{IRCNN} & {33.77} & {30.58} & {29.09} & {34.64} & {31.34} & {29.70} & {35.18} & {31.75} & {30.02} & {35.58} & {32.02} & {30.18} & {-/12.16s}\\ 
    		{ISTANet} & {32.50} & {30.78} & {29.22} & {34.84} & {31.72} & {29.85} & {35.37} & {32.04} & {30.10} & {35.87} & {32.25} & {30.22} & \bf{-/0.03s}\\ 
    		{TFPnP} & \bf{34.44} & \underline{31.23} & \underline{29.58} & \underline{35.30} & \underline{31.82} & \underline{30.11} & \underline{35.83} & \underline{32.16} & \underline{30.31} & \underline{36.19} & \underline{32.36} & \underline{30.47} & {-/0.05s}\\
    		{DPUNet} & \underline{34.39} & \bf{31.44} & \bf{29.74} & \bf{35.41} & \bf{32.11} & \bf{30.32} & \bf{35.98} & \bf{32.43} & \bf{30.59} & \bf{36.33} & \bf{32.59} & \bf{30.70} & \underline{-/0.04s}\\ 
    		\hline 
    	\end{tabular}
    	
    \end{table*}

    \begin{figure*}[t]
    	\centering
    	\setlength\tabcolsep{0.8pt}
    	\begin{tabular}{cccccccc}
    		{RecPF} & {FCSA} & {DMRI} & {IRCNN} & {ISTANet} & {TFPnP} & {DPUNet} & {GroundTruth}\\
    		\includegraphics[width=.12\linewidth,clip,keepaspectratio]{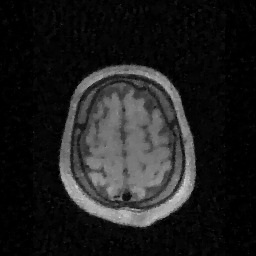} &
    		\includegraphics[width=.12\linewidth,clip,keepaspectratio]{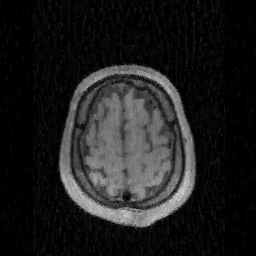} &
    		\includegraphics[width=.12\linewidth,clip,keepaspectratio]{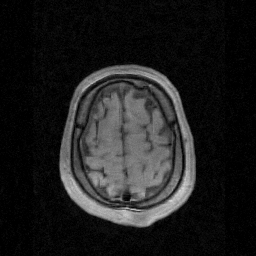} &
    		\includegraphics[width=.12\linewidth,clip,keepaspectratio]{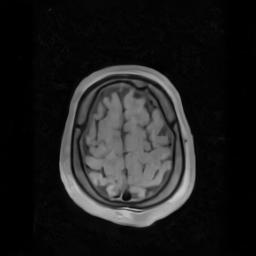} &
    		\includegraphics[width=.12\linewidth,clip,keepaspectratio]{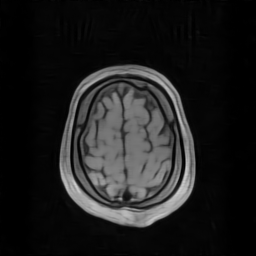} &
    		\includegraphics[width=.12\linewidth,clip,keepaspectratio]{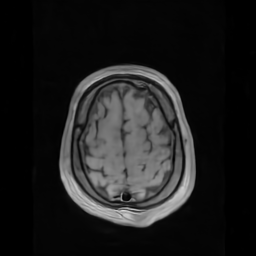} &
    		\includegraphics[width=.12\linewidth,clip,keepaspectratio]{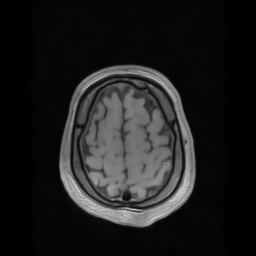} &
    		\includegraphics[width=.12\linewidth,clip,keepaspectratio]{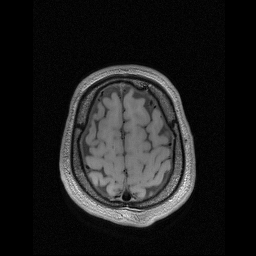} \\
    		\footnotesize  30.73 & \footnotesize 30.31 & \footnotesize  31.97 & \footnotesize 32.84 & \footnotesize 30.87 & \footnotesize 33.50 & \footnotesize 33.69 & \footnotesize  PSNR\\
    		\includegraphics[width=.12\linewidth,clip,keepaspectratio]{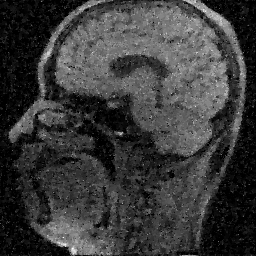} &
    		\includegraphics[width=.12\linewidth,clip,keepaspectratio]{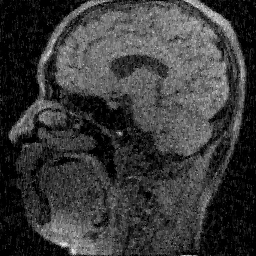} &
    		\includegraphics[width=.12\linewidth,clip,keepaspectratio]{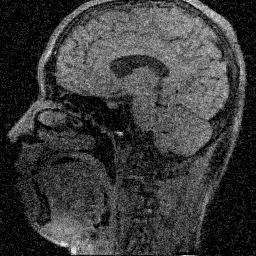} &
    		\includegraphics[width=.12\linewidth,clip,keepaspectratio]{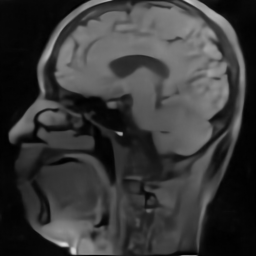} &
    		\includegraphics[width=.12\linewidth,clip,keepaspectratio]{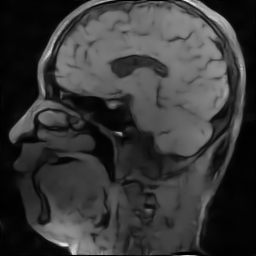} &
    		\includegraphics[width=.12\linewidth,clip,keepaspectratio]{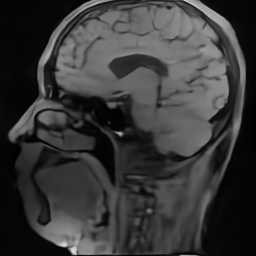} &
    		\includegraphics[width=.12\linewidth,clip,keepaspectratio]{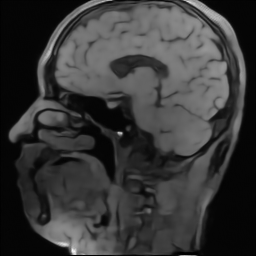} &
    		\includegraphics[width=.12\linewidth,clip,keepaspectratio]{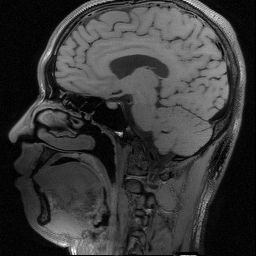} \\
    		\footnotesize  23.60 & \footnotesize 23.09 & \footnotesize  22.70 & \footnotesize 27.99 & \footnotesize  28.38 & \footnotesize  28.38 & \footnotesize 28.60 & \footnotesize  PSNR\\
    		\includegraphics[width=.12\linewidth,clip,keepaspectratio]{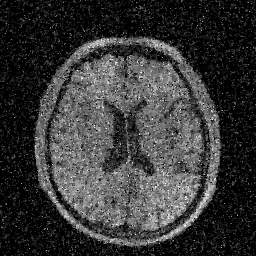} &
    		\includegraphics[width=.12\linewidth,clip,keepaspectratio]{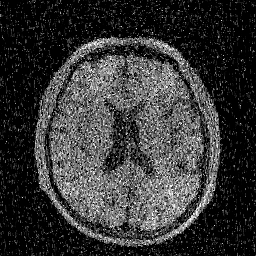} &
    		\includegraphics[width=.12\linewidth,clip,keepaspectratio]{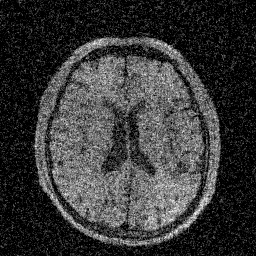} &
    		\includegraphics[width=.12\linewidth,clip,keepaspectratio]{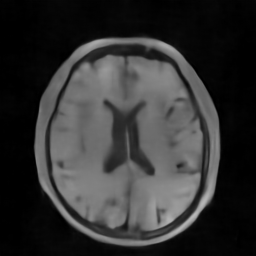} &
    		\includegraphics[width=.12\linewidth,clip,keepaspectratio]{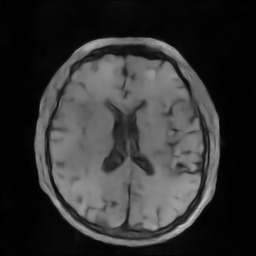} &
    		\includegraphics[width=.12\linewidth,clip,keepaspectratio]{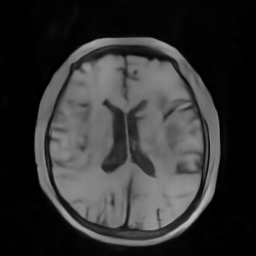} &
    		\includegraphics[width=.12\linewidth,clip,keepaspectratio]{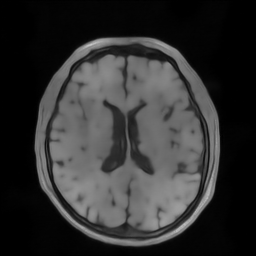} &
    		\includegraphics[width=.12\linewidth,clip,keepaspectratio]{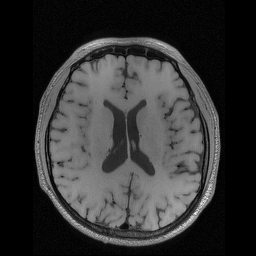} \\
    		\footnotesize  19.51 & \footnotesize 16.80 & \footnotesize  19.10 & \footnotesize 28.46 & \footnotesize  28.60 & \footnotesize 28.78 & \footnotesize 29.39 & \footnotesize  PSNR\\
    		
    	\end{tabular}
    	
    	\caption{Reconstructed images and corresponding PSNRs (dB) for CS-MRI under different sampling ratios ($\eta=20\%, 30\%, 40\%$, from top to bottom) and different noise levels ($\alpha=10, 30, 50$, from top to bottom) with seven algorithms.}
    	\label{fig:csmribenchmark}
    \end{figure*}
    
    \subsubsection{Validations on CS-MRI}
    
    We mainly compare DPUNet with six competing approaches for CS-MRI, including three classic algorithms RecPF \cite{yang2010fast}, FCSA \cite{huang2011efficient} and D-MRI \cite{eksioglu2016decoupled}, the deep unrolling network ISTANet \cite{zhang2018ista}, the state-of-the-art plug-and-play (PnP) approaches IRCNN \cite{zhang2017learning} and TFPnP \cite{wei2020tuning-free}. We train separate models of ISTANet for each sampling ratio setting in 200 epochs, and retrain TFPnP to adapt to the size of our MR images (\ie $256\times 256$ pixels). Table \ref{tab:csmribenchmark} shows the quantitative performance comparisons on 50 brain medical images under different imaging conditions. It can be seen that DPUNet significantly outperforms other competing algorithms under various imaging conditions with only one trained model. The fact that one single DPUNet model performed so well on almost all sampling ratios, particularly in comparison to the ISTA-Net models, which were trained separately per sampling ratio, again demonstrates that DPUNet generalizes across measurement matrices. What's more, compared with the plug-and-play approach IRCNN and TFPnP, DPUNet shows higher reconstruction performance owing to end-to-end training. In Fig.~\ref{fig:csmribenchmark}, we show the reconstructions of three brain MR images and corresponding PSNRs under different sampling ratios ($\eta = 20\%, 30\%, 40\%$) and noise levels ($\alpha = 10, 30, 50$) with seven algorithms. It can be observed that DPUNet can reconstruct more details and sharper edges, especially in case of severe noise.
        
    \subsubsection{Validations on CPR}
    
\begin{figure}[!t]
    	\centering
    	\setlength\tabcolsep{0.8pt}
    	\begin{tabular}{ccc}
    		{prGAMP} & {prDeep} & {DPUNet}\\
    		\includegraphics[width=.32\linewidth,clip,keepaspectratio]{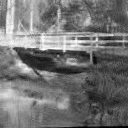} &
    		\includegraphics[width=.32\linewidth,clip,keepaspectratio]{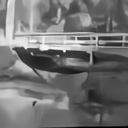} &
    		\includegraphics[width=.32\linewidth,clip,keepaspectratio]{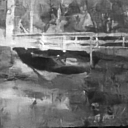} \\
    		\footnotesize 25.65 & \footnotesize 24.69 & \footnotesize 25.97 \\
    		\includegraphics[width=.32\linewidth,clip,keepaspectratio]{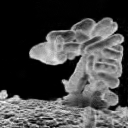} &
    		\includegraphics[width=.32\linewidth,clip,keepaspectratio]{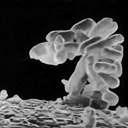} &
    		\includegraphics[width=.32\linewidth,clip,keepaspectratio]{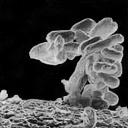} \\
    		\footnotesize 25.41 & \footnotesize 26.86 & \footnotesize 27.56 \\
    		\includegraphics[width=.32\linewidth,clip,keepaspectratio]{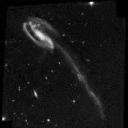} &
    		\includegraphics[width=.32\linewidth,clip,keepaspectratio]{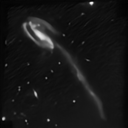} &
    		\includegraphics[width=.32\linewidth,clip,keepaspectratio]{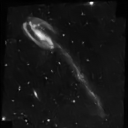} \\
    		\footnotesize 35.08 & \footnotesize 34.41 & \footnotesize 36.60 \\
    	\end{tabular}
    	\caption{Reconstructed images and corresponding PSNRs (dB) for CPR under different sampling ratios ($\eta=30\%, 40\%, 50\%$, from top to bottom) and noise level ($\alpha = 10$) with three algorithms.}
    	\label{fig:cprbenchmark}
    \end{figure}

    We mainly compare DPUNet with two state-of-the-art approaches (BM3D-prGAMP \cite{metzler2016bm3d-prgamp:} and prDeep \cite{metzler2018prdeep}) for CPR. We use their respective authors’ implementations and adopt the twelve images used in \cite{metzler2018prdeep} to quantitatively evaluate different CPR methods. The results of performance comparisons for CPR are summarized in Table~\ref{tab:cprbenchmark}. It can be seen that DPUNet can handle these imaging settings with state-of-the-art results via one single trained network. The visual comparison can be found in Fig.~\ref{fig:cprbenchmark}. It can be found that our method can still effectively recover desired images, and produce clearer results than other competing methods.
    
    \begin{table}[!t]
    	\centering
    	\small
    	\caption{Average PSNR (dB) performance comparisons for CPR on twelve images under different imaging conditions ($\eta$ and $\alpha$ denote the sampling ratio and noise level respectively). The best results are labeled in \textbf{bold} and the second-best results are \underline{underlined}.}
    	\label{tab:cprbenchmark}
    	\medskip
    	\begin{tabular}{|c|c|ccc|}
    		\hline  
    		\centering
    		{\textsc{$\eta$}} & {\textsc{$\alpha$}} & {prGAMP} & {prDeep} & {DPUNet}\\ 
    		\hline
    		\centering
    		\multirow{2}{*}{$50\%$} & {$10$} & \underline{31.56} & {30.64} & {\textbf{33.14}}\\
    		& {$30$} & {27.46} & \underline{27.69} & {\textbf{28.63}}\\
    		\hline
    		\centering
    		\multirow{2}{*}{$40\%$} & {$10$} & \underline{31.26} & {30.10} & {\textbf{32.32}}\\
    		& {$30$} & {26.77} & \underline{27.46} & {\textbf{28.31}}\\
    		\hline
    		\centering
    		\multirow{2}{*}{$30\%$} & {$10$} & {29.05} & \underline{30.00} & {\textbf{30.96}}\\
    		& {$30$} & {26.47} & \underline{26.99} & {\textbf{27.75}}\\
    		\hline
    	\end{tabular}
    	
    \end{table}
    
	\subsection{Discussion}
	
	\subsubsection{Effects of different dynamic architectures}
	
	To give some insights of the proposed dynamic proximal mapping module, we conduct an ablation study on different dynamic network architectures, including (1) PUNet: the basic proximal unrolling network without fully connected layers; (2) PUNet + DConv1: PUNet with dynamic convolution via one fully connected layer; (3) PUNet + DConv2: PUNet with dynamic convolution via two fully connected layers; (4) PUNet + DIn1: PUNet with dynamic instance normalization via one fully connected layer; (5) PUNet + DIn2: PUNet with dynamic instance normalization via two fully connected layers; (6) PUNet + DIn3: PUNet with dynamic instance normalization via three fully connected layers; (7) PUNet + DConv1In1: PUNet with dynamic convolution and instance normalization via one fully connected layer; (8) PUNet + DConv2In1: PUNet with dynamic convolution via two fully connected layers and instance normalization via one fully connected layer; (9) PUNet + DConv2In2: PUNet with dynamic convolution and instance normalization via two fully connected layers; (10) DPUNet: PUNet with dynamic convolution via one fully connected layer and dynamic instance normalization via two fully connected layers. All models are trained and tested for image CS tasks on the same experimental setting with noise-free.
	
	To compare the performance, we show PSNRs (dB) of BCS reconstructions ($\eta=10\%, 25\%, 40\%$) on Set11 with different structures and the number of model parameters (Million), provided in Table~\ref{tab:ablation}. 
	It can be seen that the performance of PUNet could be significantly improved by using dynamic convolution or dynamic instance normalization with two fully connected layers, especially by using them together. Considered by the complexity of the model, our final choice (\ie DPUNet) achieves performance near to the top with a reasonable number of parameters.
	
	\begin{table}[t]
		\centering
		\small
		\caption{Comparisons of different structures in the proposed dynamic proximal mapping module. We show PSNRs (dB) of BCS reconstructions under multiple CS ratios on Set11 with different architectures and the number of model parameters (Million). The best results are labeled in \textbf{bold} and the second-best results are \underline{underlined}.
		}
		\label{tab:ablation}
		\medskip
		\begin{tabular}{|l|ccc|l|}
			\hline  
			\centering 
			\multirow{2}{*}{\textsc{Architectures}} & \multicolumn{3}{c|}{\textsc{CS Ratios}} & \multirow{2}{*}{\textsc{Params}} \\ \cline{2-4}
			~ & {10\%} & {25\%} & {40\%} & ~ \\  
			\hline  
			\centering   
			{PUNet} & {26.57} & {31.80} & {34.85} & {1.12} \\ 
			{PUNet + DConv1} & {26.87} & {32.47} & {35.50} & {2.24} \\ 
			{PUNet + DConv2} & {27.05} & {32.73} & {36.05} & {72.65} \\ 
			{PUNet + DIn1} & {26.87} & {32.39} & {35.70} & {1.13} \\ 
			{PUNet + DIn2} & {27.15} & {32.74} & {36.12} & {1.46} \\ 
			{PUNet + DIn3} & {27.25} & {32.85} & {36.23} & {1.79} \\ 
			{PUNet + DConv1In1} & {26.49} & {32.04} & {34.84} & {2.25} \\ 
			{PUNet + DConv2In1} & {26.76} & {32.51} & {35.92} & {72.66} \\ 
			{PUNet + DConv2In2} & \bf{27.33} & \bf{32.99} & \underline{36.42} & {72.98} \\ 
			{DPUNet} & \underline{27.31} & \underline{32.99} & \bf{36.48} & {2.58} \\
			\hline 
		\end{tabular}
		
	\end{table}
	
	\subsubsection{Effects of different optimization frameworks}
	
	To provide the insight into the choice of optimization framework, we compare the performance of the proposed three unrolling frameworks on image CS tasks, including their unrolled networks and corresponding dynamic versions.
	In Table~\ref{tab:extensions}, it can be seen that the proposed dynamic proximal mapping module can consistently boost the performance of unrolling network derived from PGD, HQS, and ADMM, with average performance gain 1.45dB, 1.60dB, and 1.45dB, respectively. Meanwhile, DPUNet-PGD achieves the highest performance under various sampling ratios compared against other networks, chosen as the final choice. 
	
	\begin{table}[http]
		\centering
		\caption{Comparisons of unrolling networks based on PGD, HQS, and ADMM. We show PSNRs (dB) of image compressive sensing reconstructions under various sampling ratios on Set11. 
		}
		\label{tab:extensions}
		\medskip
		\begin{tabular}{|l|cccc|}
			\hline  
			\centering
			\multirow{2}{*}{\textsc{Extensions}} & \multicolumn{4}{c|}{\textsc{Sampling Ratios}}\\ \cline{2-5}
			~ & {10\%} & {25\%} & {40\%} & {50\%}\\  
			\hline  
			\centering   
			PUNet-PGD & {26.57} & {31.80} & {34.85} & {36.12} \\ 
			DPUNet-PGD & 27.31 & 32.99 & 36.48 & 38.35 \\ 
			\hline
			PUNet-HQS & {26.05} & {31.18} & {34.16}  & {35.55}\\ 
			DPUNet-HQS & 26.74 & 32.49 & 36.08 & 38.04 \\
			\hline
			PUNet-ADMM & {26.16} & {31.30} & {34.33} & {35.80} \\ 
			DPUNet-ADMM & 26.80 & 32.51 & 36.05 & 38.02 \\ 
			\hline 
		\end{tabular}
		
	\end{table}
	
	\subsubsection{Generalizability of DPUNet}
	
\begin{figure}[http]
		\centering
		\setlength\tabcolsep{0.8pt}
		\includegraphics[width=.9\linewidth,clip,keepaspectratio]{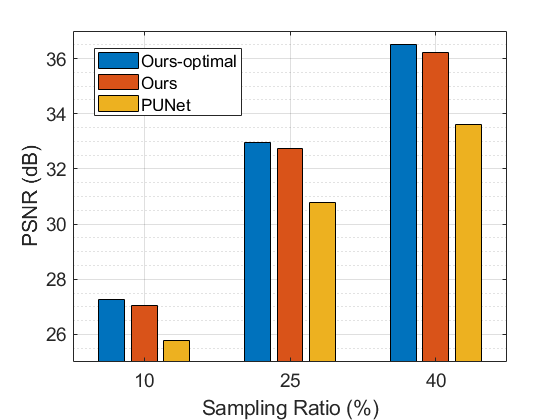}\\
		\caption{The model generalizability test of our method for image compressive sensing under multiple sampling ratios. "Ours-optimal" denotes the (three) models trained and tested on consistent sampling ratios. "Ours" is the single model trained and tested on all sampling ratios. "PUNet" is the degraded version of "Ours" without a dynamic module, and still the single model trained and tested on all sampling ratios.}
		\label{fig:generaliability}
	\end{figure}

	To further investigate the generalizability of DPUNet, we train and test DPUNet on consistent CS ratios, denoted by DPUNet-optimal. Fig.~\ref{fig:generaliability} shows the comparison between the "DPUNet-optimal", "DPUNet", and "PUNet" models for image compressive sensing under multiple CS ratios. Note that "DPUNet-optimal" denotes the (three) models of DPUNet separately trained and tested on each CS ratio, which is expected to get the optimal results. "DPUNet" and "PUNet" indicate the single model trained and tested on all CS ratios, and the main difference between them is that "DPUNet" adopts the proposed dynamic module which "PUNet" lacks.
	
	It can be observed that a single trained model of DPUNet can achieve very close performance (the average PSNR difference is about 0.24 dB) with the optimal results, while there is a large gap (the average PSNR difference is about 2.18 dB) between "PUNet" model and the optimal model. 
	Overall, we demonstrate that the proposed dynamic proximal mapping module can significantly boost the generalizability of deep unrolling networks, avoiding time-consuming and storage-consuming retraining.
	
	\subsubsection{Robustness on imaging parameters mismatch}
	
\begin{figure}[!t]
		\centering
		\setlength\tabcolsep{0.8pt}
		\includegraphics[width=.9\linewidth,clip,keepaspectratio]{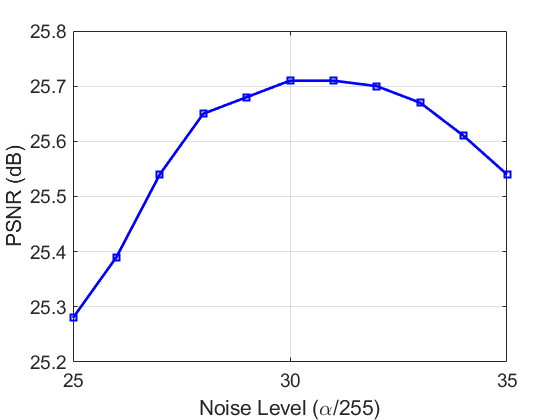}\\
		\caption{The model robustness test of DPUNet with mismatched imaging parameters on noisy BCS measurements. The real imaging conditions are sampling ratio $\eta=30\%$, noise level $\alpha=30$. Here we test DPUNet under a range of parameter values of the noise level as the input and show corresponding reconstruction results (PSNR) on Set11.}
		\label{fig:robustness}
	\end{figure}

\begin{figure}[!t]
		\centering
		\setlength\tabcolsep{0.8pt}
		\begin{tabular}{ccc}
			{$\alpha=25$} & {$\alpha=28$} & {$\alpha=30$*}\\
			\includegraphics[width=.32\linewidth,clip,keepaspectratio]{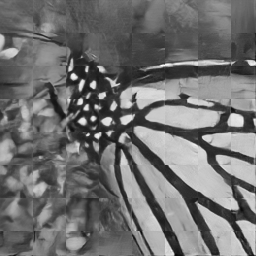} &
			\includegraphics[width=.32\linewidth,clip,keepaspectratio]{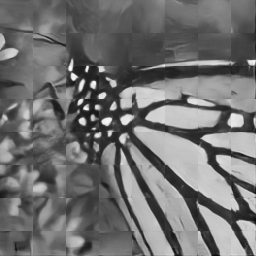} &
			\includegraphics[width=.32\linewidth,clip,keepaspectratio]{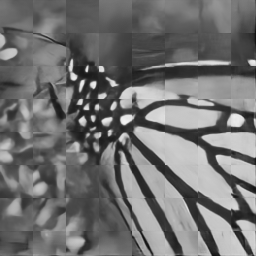} \\
			\footnotesize 24.81 & \footnotesize 25.11 & \footnotesize 25.11 \\
			{$\alpha=33$} & {$\alpha=35$} & {GroundTruth}\\
			\includegraphics[width=.32\linewidth,clip,keepaspectratio]{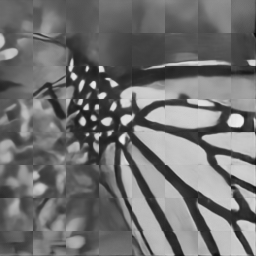} &
			\includegraphics[width=.32\linewidth,clip,keepaspectratio]{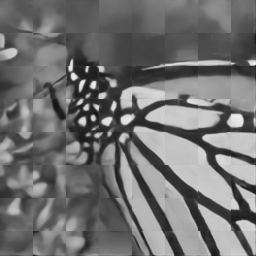} &
			\includegraphics[width=.32\linewidth,clip,keepaspectratio]{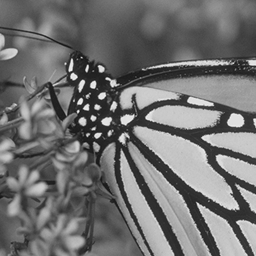} \\
			\footnotesize 25.00 & \footnotesize 25.01 & \footnotesize PSNR \\
		\end{tabular}
		\caption{Reconstructed images and corresponding PSNRs (dB) from noisy BCS measurements ($\eta=30\%, \alpha=30$) under a range of parameter values of the noise-level at the inference stage.}
		\label{fig:visualcomparisons}
	\end{figure}

	In practical applications, while the sampling ratio can be accurately measured, the noise level or signal-to-noise ratio (SNR) is often estimated by related methods~\cite{pauluzzi2000comparison,riffe2007snr,dietrich2007measurement} since the ground truth is unknown. Thereby, DPUNet should be robust to mismatched imaging parameters, mainly under noisy conditions. To analyze the robustness of DPUNet for the mismatched imaging parameters, we use a range of parameter values of the noise-level as the input at the inference stage, and show average reconstruction results (PSNRs) on Set11 test set from noisy BCS measurements (sampling ratio $\eta=30\%$, noise level $\alpha=30$). Fig.~\ref{fig:robustness} illuminates that DPUNet is robust to the inaccuracy of imaging parameters -- reaching similar reconstruction results under a range of mismatched imaging parameters. The visual comparisons can see Fig.~\ref{fig:visualcomparisons}, which also shows  visually similar results of reconstructed images.
	
	\subsubsection{Hyper-parameters analysis}
	
\begin{figure}[!t]
		\centering
		\setlength\tabcolsep{0.8pt}
		\includegraphics[width=.9\linewidth,clip,keepaspectratio]{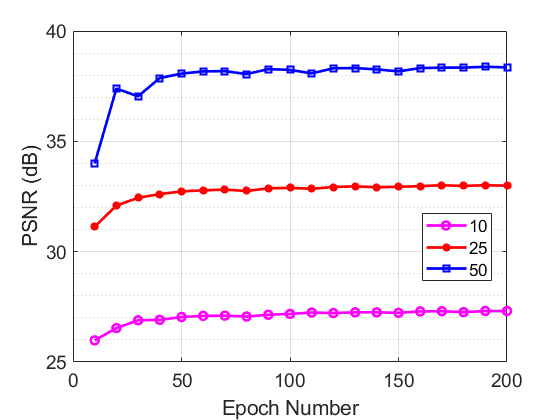}
		
		\caption{The model performance of BCS reconstructions ($\eta=10\%, 25\%, 50\%$) on Set11 under different training epochs.}
		\label{fig:epoch}
	\end{figure}

	We conduct a hyper-parameters analysis and train different models of our method, then present experimental results. Specifically, we focus on two hyper-parameters of our DPUNet, \ie unrolling iterations and training epochs. First, we unroll different numbers of proximal optimization iterations and train corresponding models. The performance of different models on image compressive sensing tasks is presented in Table~\ref{tab:unrolling}. It can be seen that unrolling ten iterations is sufficient for our DPUNet to get promising results.
	Furthermore, We provide the performance of trained models on different epochs on the image compressive sensing task in Fig.~\ref{fig:epoch}. It can be observed that our model trained by 200 epochs can achieve satisfactory performance.
	
	\begin{table}[!t]
		\centering
		\small
		\caption{Average PSNR (dB) performance comparisons of DPUNet with different unrolling iterations. The best results are labeled in \textbf{bold} and the second-best results are \underline{underlined}.}
		\label{tab:unrolling}
		
		\begin{tabular}{|c|ccccc|}
			\hline  
			\centering 
			\multirow{2}{*}{\textsc{Model}} & \multicolumn{5}{c|}{\textsc{CS Ratio (on Set11)}}\\
			~ & {4\%} & {10\%} & {25\%} & {40\%} & {50\%}\\  
			\hline  
			\centering   
			{DPUNet-5} & {21.56} & {26.09} & {31.08} & {35.00} & {36.88}\\ 
			{DPUNet-10} & \underline{22.09} & \underline{27.04} & \underline{32.73} & \underline{36.22} & \underline{38.07}\\ 
			{DPUNet-15} & \bf{22.14} & \bf{27.08} & \bf{32.76} & \bf{36.23} & \bf{38.13}\\ 
			\hline 
		\end{tabular}
		
	\end{table}
	
	\subsection{Extension for multiple compressive imaging tasks}
	
	Inspired by the same network structure for different compressive imaging modalities, we explore the potential of our method to handle multiple imaging tasks with one single trained model simultaneously. To verify this claim, we present an extension of our method and run a mixed experiment, where we consider three imaging modalities mentioned above under different sampling ratios with varying noise levels. To implement this, we simulate to generate the BCS measurements with the sampling ratio $\eta$ uniformly sampled from \{0.01,0.04,0.1,0.25,0.4,0.5\} from CS training data-set, CS-MRI measurements with the sampling ratio $\eta$ uniformly sampled from \{0.2,0.3,0.4,0.5\} from MRI training data-set, and the under-sampling coded diffraction measurements uniformly sampled from \{0.3,0.4,0.5\} from CPR training data-set, all adding noise with $\alpha$ uniformly sampled from [0,50]. 
	
	Moreover, we add an extra parameter $\kappa$ to represent the imaging modality~\footnote{For simplicity, we input the parameters $\kappa = 1$ for dealing with BCS task, $\kappa = 2$ for CS-MRI, and $\kappa = 3$ for CPR.}, and the other parameters are related to the imaging condition ($\eta, \alpha$). During training, the BCS, CS-MRI, and CPR training data pairs that include image patches and corresponding measurements are alternately fed to our network, together with imaging parameters \{$\kappa, \eta, \alpha$\}. The network is trained in 200000 iterations using pixel-wise $L_2$ loss and Adam optimizer with learning rate $10^{-4}$. It takes about 18 hours to train the model.
	
	The reconstruction results with the extension of our method for all imaging modalities and imaging conditions are provided in Table~\ref{tab:Extension}. We can see that our method can not only handle various imaging conditions but also totally different imaging modalities via one single model without retraining. Here we also compare with the results of our baselines, which represent the models of DPUNet trained on a single imaging task (BCS, CS-MRI, or CPR), respectively. It can be seen that the extension of our method still achieves close results to the baselines, despite small performance degradation. It means with the same network structure, our method can handle different imaging tasks via one single trained model without losing much accuracy. We show reconstructed results of a single trained model for multiple imaging modalities with noise level ($\alpha = 10$) and different sampling ratios ($\eta = 10\%$ for BCS, $\eta = 20\%$ for CS-MRI, $\eta = 50\%$ for CPR) in Fig. \ref{fig:review_results}. It can be found that the proposed network is flexible for various imaging conditions and universal for different imaging modalities.
	
	\begin{table}[htbp]
		\centering
		\small
		\caption{Average PSNRs (dB) of reconstruction results with our single trained model (Ours*) for different imaging modalities under varying imaging conditions. The baselines are the models of DPUNet separately trained on a single imaging task (BCS, CS-MRI, and CPR), which represent the performance standard. }
		\label{tab:Extension}
		\medskip
		\begin{tabular}{|c|c|c|c|c|}
			\hline  
			\centering 
			{Task} & {$\eta$} & {$\alpha$} & {Ours*} & {Baselines}\\
			\hline
			\multirow{4}{*}{\textsc{BCS}} & \multirow{2}{*}{$30\%$} & {$10$} & {29.96} & {30.04}\\ 
			\centering   
			~ & ~ & {$30$} & {25.57} & {25.70}\\ 
			\cline{2-5} 
			~ & \multirow{2}{*}{$50\%$} & {$10$} & {31.89} & {31.99} \\
			\centering   
			~ & ~ & {$30$} & {27.03} & {27.20}\\ 
			\hline 
			\multirow{4}{*}{\textsc{CS-MRI}} & \multirow{2}{*}{$30\%$} & {$10$} & {35.08} & {35.41}\\ 
			\centering   
			~ & ~ & {$30$} & {31.61} & {32.08}\\ 
			\cline{2-5} 
			~ & \multirow{2}{*}{$50\%$} & {$10$} & {35.97} & {36.33} \\
			\centering   
			~ & ~ & {$30$} & {32.07} & {32.59}\\ 
			\hline 
			\multirow{4}{*}{\textsc{CPR}} & \multirow{2}{*}{$30\%$} & {$10$} & {30.56} & {30.96}\\ 
			\centering   
			~ & ~ & {$30$} & {27.75} & {27.75}\\ 
			\cline{2-5} 
			~ & \multirow{2}{*}{$50\%$} & {$10$} & {32.91} & {33.14} \\
			\centering   
			~ & ~ & {$30$} & {28.38} & {28.63}\\ 
			\hline 
		\end{tabular}
		
	\end{table}
	
	\section{Conclusion}
	\label{sec:conclusion}
	
	We have proposed a dynamic proximal unrolling network for a variety of compressive imaging problems under varying imaging conditions. The main contribution of the proposed method is developing a dynamic proximal mapping module, which can dynamically update parameters of the proximal network at the inference stage and make it adapt to different imaging settings and even imaging tasks. As a result, the proposed method can handle a wide range of compressive imaging modalities, including image compressive sensing, CS-MRI, and compressive phase retrieval under varying imaging conditions via one single trained model. Experimental results demonstrate the effectiveness and state-of-the-art performance of the proposed method. Thereby, we envision the proposed network to be applied to embedded mobile devices where storage and computational resources demands become prohibitive, and to handle a variety of imaging tasks via only one trained model.
	
	\bibliographystyle{IEEEtran}
	\bibliography{bibfile}

\begin{thebibliography}{10}
\providecommand{\url}[1]{#1}
\csname url@rmstyle\endcsname
\providecommand{\newblock}{\relax}
\providecommand{\bibinfo}[2]{#2}
\providecommand\BIBentrySTDinterwordspacing{\spaceskip=0pt\relax}
\providecommand\BIBentryALTinterwordstretchfactor{4}
\providecommand\BIBentryALTinterwordspacing{\spaceskip=\fontdimen2\font plus
\BIBentryALTinterwordstretchfactor\fontdimen3\font minus
  \fontdimen4\font\relax}
\providecommand\BIBforeignlanguage[2]{{%
\expandafter\ifx\csname l@#1\endcsname\relax
\typeout{** WARNING: IEEEtran.bst: No hyphenation pattern has been}%
\typeout{** loaded for the language `#1'. Using the pattern for}%
\typeout{** the default language instead.}%
\else
\language=\csname l@#1\endcsname
\fi
#2}}

\bibitem{donoho2006compressed}
D.~L. Donoho, ``Compressed sensing,'' \emph{IEEE Transactions on information
  theory}, vol.~52, no.~4, pp. 1289--1306, 2006.

\bibitem{liutkus2014imaging}
A.~Liutkus, D.~Martina, S.~Popoff, G.~Chardon, O.~Katz, G.~Lerosey, S.~Gigan,
  L.~Daudet, and I.~Carron, ``Imaging with nature: Compressive imaging using a
  multiply scattering medium,'' \emph{Scientific reports}, vol.~4, no.~1, pp.
  1--7, 2014.

\bibitem{duarte2008single}
M.~F. Duarte, M.~A. Davenport, D.~Takhar, J.~N. Laska, T.~Sun, K.~F. Kelly, and
  R.~G. Baraniuk, ``Single-pixel imaging via compressive sampling,'' \emph{IEEE
  signal processing magazine}, vol.~25, no.~2, pp. 83--91, 2008.

\bibitem{kerviche2014information}
R.~Kerviche, N.~Zhu, and A.~Ashok, ``Information-optimal scalable compressive
  imaging system,'' in \emph{Computational Optical Sensing and Imaging}.\hskip
  1em plus 0.5em minus 0.4em\relax Optical Society of America, 2014, pp.
  CM2D--2.

\bibitem{sankaranarayanan2012cs}
A.~C. Sankaranarayanan, C.~Studer, and R.~G. Baraniuk, ``Cs-muvi: Video
  compressive sensing for spatial-multiplexing cameras,'' in \emph{2012 IEEE
  International Conference on Computational Photography (ICCP)}.\hskip 1em plus
  0.5em minus 0.4em\relax IEEE, 2012, pp. 1--10.

\bibitem{lustig2007sparse}
M.~Lustig, D.~Donoho, and J.~M. Pauly, ``Sparse mri: The application of
  compressed sensing for rapid mr imaging,'' \emph{Magnetic Resonance in
  Medicine: An Official Journal of the International Society for Magnetic
  Resonance in Medicine}, vol.~58, no.~6, pp. 1182--1195, 2007.

\bibitem{lustig2008compressed}
M.~Lustig, D.~L. Donoho, J.~M. Santos, and J.~M. Pauly, ``Compressed sensing
  mri,'' \emph{IEEE signal processing magazine}, vol.~25, no.~2, pp. 72--82,
  2008.

\bibitem{rousset2016adaptive}
F.~Rousset, N.~Ducros, A.~Farina, G.~Valentini, C.~D’Andrea, and F.~Peyrin,
  ``Adaptive basis scan by wavelet prediction for single-pixel imaging,''
  \emph{IEEE Transactions on Computational Imaging}, vol.~3, no.~1, pp. 36--46,
  2016.

\bibitem{yuan2021snapshot}
X.~Yuan, D.~J. Brady, and A.~K. Katsaggelos, ``Snapshot compressive imaging:
  Theory, algorithms, and applications,'' \emph{IEEE Signal Processing
  Magazine}, vol.~38, no.~2, pp. 65--88, 2021.

\bibitem{moravec2007compressive}
M.~L. Moravec, J.~K. Romberg, and R.~G. Baraniuk, ``Compressive phase
  retrieval,'' in \emph{Wavelets XII}, vol. 6701.\hskip 1em plus 0.5em minus
  0.4em\relax International Society for Optics and Photonics, 2007, p. 670120.

\bibitem{ohlsson2012cprl}
H.~Ohlsson, A.~Yang, R.~Dong, and S.~Sastry, ``Cprl--an extension of
  compressive sensing to the phase retrieval problem,'' \emph{Advances in
  Neural Information Processing Systems}, vol.~25, pp. 1367--1375, 2012.

\bibitem{ma2008efficient}
S.~Ma, W.~Yin, Y.~Zhang, and A.~Chakraborty, ``An efficient algorithm for
  compressed mr imaging using total variation and wavelets,'' in \emph{IEEE
  Conference on Computer Vision and Pattern Recognition (CVPR)}.\hskip 1em plus
  0.5em minus 0.4em\relax IEEE, 2008, pp. 1--8.

\bibitem{liao2008sparse}
H.~Y. Liao and G.~Sapiro, ``Sparse representations for limited data
  tomography,'' in \emph{IEEE International Symposium on Biomedical Imaging:
  From Nano to Macro}.\hskip 1em plus 0.5em minus 0.4em\relax IEEE, 2008, pp.
  1375--1378.

\bibitem{dong2014compressive}
W.~Dong, G.~Shi, X.~Li, Y.~Ma, and F.~Huang, ``Compressive sensing via nonlocal
  low-rank regularization,'' \emph{IEEE Transactions on Image Processing},
  vol.~23, no.~8, pp. 3618--3632, 2014.

\bibitem{qu2014magnetic}
X.~Qu, Y.~Hou, F.~Lam, D.~Guo, J.~Zhong, and Z.~Chen, ``Magnetic resonance
  image reconstruction from undersampled measurements using a patch-based
  nonlocal operator,'' \emph{Medical Image Analysis}, vol.~18, no.~6, pp.
  843--856, 2014.

\bibitem{zhang2014group}
J.~Zhang, D.~Zhao, and W.~Gao, ``Group-based sparse representation for image
  restoration,'' \emph{IEEE Transactions on Image Processing}, vol.~23, no.~8,
  pp. 3336--3351, 2014.

\bibitem{6873741}
C.~Guo and Q.~Yang, ``A neurodynamic optimization method for recovery of
  compressive sensed signals with globally converged solution approximating to
  $ l\_ $\{$0$\}$ $ minimization,'' \emph{IEEE Transactions on Neural Networks
  and Learning Systems}, vol.~26, no.~7, pp. 1363--1374, 2015.

\bibitem{metzler2016denoising}
C.~A. Metzler, A.~Maleki, and R.~G. Baraniuk, ``From denoising to compressed
  sensing,'' \emph{IEEE Transactions on Information Theory}, vol.~62, no.~9,
  pp. 5117--5144, 2016.

\bibitem{combettes2011proximal}
P.~L. Combettes and J.-C. Pesquet, ``Proximal splitting methods in signal
  processing,'' in \emph{Fixed-point algorithms for inverse problems in science
  and engineering}.\hskip 1em plus 0.5em minus 0.4em\relax Springer, 2011, pp.
  185--212.

\bibitem{afonso2010fast}
M.~V. Afonso, J.~M. Bioucas-Dias, and M.~A. Figueiredo, ``Fast image recovery
  using variable splitting and constrained optimization,'' \emph{IEEE
  transactions on image processing}, vol.~19, no.~9, pp. 2345--2356, 2010.

\bibitem{beck2009fast}
A.~Beck and M.~Teboulle, ``A fast iterative shrinkage-thresholding algorithm
  for linear inverse problems,'' \emph{SIAM Journal on Imaging Sciences},
  vol.~2, no.~1, pp. 183--202, 2009.

\bibitem{geman1995nonlinear}
D.~Geman and C.~Yang, ``Nonlinear image recovery with half-quadratic
  regularization,'' \emph{IEEE transactions on Image Processing}, vol.~4,
  no.~7, pp. 932--946, 1995.

\bibitem{boyd2011distributed}
S.~Boyd, N.~Parikh, E.~Chu, B.~Peleato, J.~Eckstein, \emph{et~al.},
  ``Distributed optimization and statistical learning via the alternating
  direction method of multipliers,'' \emph{Foundations and
  Trends{\textregistered} in Machine learning}, vol.~3, no.~1, pp. 1--122,
  2011.

\bibitem{yang2010fast}
J.~Yang, Y.~Zhang, and W.~Yin, ``A fast alternating direction method for
  tvl1-l2 signal reconstruction from partial fourier data,'' \emph{IEEE Journal
  of Selected Topics in Signal Processing}, vol.~4, no.~2, pp. 288--297, 2010.

\bibitem{ongie2020deep}
G.~Ongie, A.~Jalal, C.~A. Metzler, R.~G. Baraniuk, A.~G. Dimakis, and
  R.~Willett, ``Deep learning techniques for inverse problems in imaging,''
  \emph{IEEE Journal on Selected Areas in Information Theory}, vol.~1, no.~1,
  pp. 39--56, 2020.

\bibitem{mousavi2015deep}
A.~Mousavi, A.~B. Patel, and R.~G. Baraniuk, ``A deep learning approach to
  structured signal recovery,'' in \emph{2015 53rd annual allerton conference
  on communication, control, and computing (Allerton)}.\hskip 1em plus 0.5em
  minus 0.4em\relax IEEE, 2015, pp. 1336--1343.

\bibitem{iliadis2018deep}
M.~Iliadis, L.~Spinoulas, and A.~K. Katsaggelos, ``Deep fully-connected
  networks for video compressive sensing,'' \emph{Digital Signal Processing},
  vol.~72, pp. 9--18, 2018.

\bibitem{kulkarni2016reconnet}
K.~Kulkarni, S.~Lohit, P.~Turaga, R.~Kerviche, and A.~Ashok, ``Reconnet:
  Non-iterative reconstruction of images from compressively sensed
  measurements,'' in \emph{Proceedings of the IEEE Conference on Computer
  Vision and Pattern Recognition}, 2016, pp. 449--458.

\bibitem{gregor2010learning}
K.~Gregor and Y.~LeCun, ``Learning fast approximations of sparse coding,'' in
  \emph{Proceedings of the 27th international conference on international
  conference on machine learning}, 2010, pp. 399--406.

\bibitem{borgerding2017amp}
M.~Borgerding, P.~Schniter, and S.~Rangan, ``Amp-inspired deep networks for
  sparse linear inverse problems,'' \emph{IEEE Transactions on Signal
  Processing}, vol.~65, no.~16, pp. 4293--4308, 2017.

\bibitem{yang2016deep}
Y.~Yang, J.~Sun, H.~Li, and Z.~Xu, ``Deep admm-net for compressive sensing
  mri,'' in \emph{Proceedings of the 30th international conference on neural
  information processing systems}, 2016, pp. 10--18.

\bibitem{zhang2018ista}
J.~Zhang and B.~Ghanem, ``Ista-net: Interpretable optimization-inspired deep
  network for image compressive sensing,'' in \emph{Proceedings of the IEEE
  conference on computer vision and pattern recognition}, 2018, pp. 1828--1837.

\bibitem{mun2009block}
S.~Mun and J.~E. Fowler, ``Block compressed sensing of images using directional
  transforms,'' in \emph{2009 16th IEEE international conference on image
  processing (ICIP)}.\hskip 1em plus 0.5em minus 0.4em\relax IEEE, 2009, pp.
  3021--3024.

\bibitem{ravishankar2010mr}
S.~Ravishankar and Y.~Bresler, ``Mr image reconstruction from highly
  undersampled k-space data by dictionary learning,'' \emph{IEEE Transactions
  on Medical Imaging}, vol.~30, no.~5, pp. 1028--1041, 2010.

\bibitem{9361634}
Z.~Zha, B.~Wen, X.~Yuan, J.~Zhou, C.~Zhu, and A.~C. Kot, ``A hybrid structural
  sparsification error model for image restoration,'' \emph{IEEE Transactions
  on Neural Networks and Learning Systems}, pp. 1--15, 2021.

\bibitem{mairal2009non}
J.~Mairal, F.~R. Bach, J.~Ponce, G.~Sapiro, and A.~Zisserman, ``Non-local
  sparse models for image restoration,'' in \emph{The IEEE International
  Conference on Computer Vision (ICCV)}, vol.~29, 2009, pp. 54--62.

\bibitem{mousavi2017learning}
A.~Mousavi and R.~G. Baraniuk, ``Learning to invert: Signal recovery via deep
  convolutional networks,'' in \emph{IEEE International Conference on
  Acoustics, Speech and Signal Processing (ICASSP)}.\hskip 1em plus 0.5em minus
  0.4em\relax IEEE, 2017, pp. 2272--2276.

\bibitem{adler2016deep}
A.~Adler, D.~Boublil, M.~Elad, and M.~Zibulevsky, ``A deep learning approach to
  block-based compressed sensing of images,'' \emph{arXiv preprint
  arXiv:1606.01519}, 2016.

\bibitem{9484349}
M.~Yamaç, M.~Ahishali, S.~Kiranyaz, and M.~Gabbouj, ``Convolutional sparse
  support estimator network (csen): From energy-efficient support estimation to
  learning-aided compressive sensing,'' \emph{IEEE Transactions on Neural
  Networks and Learning Systems}, pp. 1--15, 2021.

\bibitem{9470780}
C.-M. Feng, Z.~Yang, H.~Fu, Y.~Xu, J.~Yang, and L.~Shao, ``Donet: Dual-octave
  network for fast mr image reconstruction,'' \emph{IEEE Transactions on Neural
  Networks and Learning Systems}, pp. 1--11, 2021.

\bibitem{shi2019scalable}
W.~Shi, F.~Jiang, S.~Liu, and D.~Zhao, ``Scalable convolutional neural network
  for image compressed sensing,'' in \emph{Proceedings of the IEEE/CVF
  Conference on Computer Vision and Pattern Recognition}, 2019, pp.
  12\,290--12\,299.

\bibitem{hershey2014deep}
J.~R. Hershey, J.~L. Roux, and F.~Weninger, ``Deep unfolding: Model-based
  inspiration of novel deep architectures,'' \emph{arXiv preprint
  arXiv:1409.2574}, 2014.

\bibitem{sun2016deep}
J.~Sun, H.~Li, Z.~Xu, \emph{et~al.}, ``Deep admm-net for compressive sensing
  mri,'' in \emph{Advances in neural information processing systems}, 2016, pp.
  10--18.

\bibitem{NIPS2017_6774}
C.~Metzler, A.~Mousavi, and R.~Baraniuk, ``Learned d-amp: Principled neural
  network based compressive image recovery,'' in \emph{Advances in Neural
  Information Processing Systems}, 2017, pp. 1772--1783.

\bibitem{aggarwal2018modl}
H.~K. Aggarwal, M.~P. Mani, and M.~Jacob, ``Modl: Model-based deep learning
  architecture for inverse problems,'' \emph{IEEE transactions on medical
  imaging}, vol.~38, no.~2, pp. 394--405, 2018.

\bibitem{dong2018denoising}
W.~Dong, P.~Wang, W.~Yin, G.~Shi, F.~Wu, and X.~Lu, ``Denoising prior driven
  deep neural network for image restoration,'' \emph{IEEE Transactions on
  Pattern Analysis and Machine Intelligence}, vol.~41, no.~10, pp. 2305--2318,
  2018.

\bibitem{wang2019deep}
L.~Wang, C.~Sun, Y.~Fu, M.~H. Kim, and H.~Huang, ``Hyperspectral image
  reconstruction using a deep spatial-spectral prior,'' in \emph{Proceedings of
  the IEEE/CVF Conference on Computer Vision and Pattern Recognition}, 2019,
  pp. 8032--8041.

\bibitem{zhang2020deep}
K.~Zhang, L.~V. Gool, and R.~Timofte, ``Deep unfolding network for image
  super-resolution,'' in \emph{Proceedings of the IEEE/CVF Conference on
  Computer Vision and Pattern Recognition}, 2020, pp. 3217--3226.

\bibitem{venkatakrishnan2013plug}
S.~V. Venkatakrishnan, C.~A. Bouman, and B.~Wohlberg, ``Plug-and-play priors
  for model based reconstruction,'' in \emph{IEEE Global Conference on Signal
  and Information Processing}.\hskip 1em plus 0.5em minus 0.4em\relax IEEE,
  2013, pp. 945--948.

\bibitem{Chang_2017_ICCV}
J.~Rick~Chang, C.-L. Li, B.~Poczos, B.~Vijaya~Kumar, and A.~C.
  Sankaranarayanan, ``One network to solve them all--solving linear inverse
  problems using deep projection models,'' in \emph{Proceedings of the IEEE
  International Conference on Computer Vision}, 2017, pp. 5888--5897.

\bibitem{metzler2018prdeep}
C.~A. Metzler, P.~Schniter, A.~Veeraraghavan, and R.~G. Baraniuk, ``prdeep:
  Robust phase retrieval with a flexible deep network,'' in \emph{international
  conference on machine learning}, 2018, pp. 3498--3507.

\bibitem{Meinhardt_2017_ICCV}
T.~Meinhardt, M.~Moller, C.~Hazirbas, and D.~Cremers, ``Learning proximal
  operators: Using denoising networks for regularizing inverse imaging
  problems,'' in \emph{Proceedings of the IEEE International Conference on
  Computer Vision}, 2017, pp. 1781--1790.

\bibitem{wei2020tuning-free}
K.~Wei, A.~Aviles-Rivero, J.~Liang, Y.~Fu, C.-B. Sch{\"o}nlieb, and H.~Huang,
  ``Tuning-free plug-and-play proximal algorithm for inverse imaging
  problems,'' in \emph{International Conference on Machine Learning}.\hskip 1em
  plus 0.5em minus 0.4em\relax PMLR, 2020, pp. 10\,158--10\,169.

\bibitem{zhang2017learning}
K.~Zhang, W.~Zuo, S.~Gu, and L.~Zhang, ``Learning deep cnn denoiser prior for
  image restoration,'' in \emph{Proceedings of the IEEE conference on computer
  vision and pattern recognition}, 2017, pp. 3929--3938.

\bibitem{metzler2016bm3d-prgamp:}
C.~A. Metzler, A.~Maleki, and R.~G. Baraniuk, ``Bm3d-prgamp: Compressive phase
  retrieval based on bm3d denoising,'' in \emph{2016 IEEE International
  Conference on Image Processing (ICIP)}.\hskip 1em plus 0.5em minus
  0.4em\relax IEEE, 2016, pp. 2504--2508.

\bibitem{wei2020tfpnp}
K.~Wei, A.~Aviles-Rivero, J.~Liang, Y.~Fu, H.~Huang, and C.-B. Sch{\"o}nlieb,
  ``Tfpnp: Tuning-free plug-and-play proximal algorithm with applications to
  inverse imaging problems,'' \emph{arXiv preprint arXiv:2012.05703}, 2020.

\bibitem{ulyanov2017improved}
D.~Ulyanov, A.~Vedaldi, and V.~Lempitsky, ``Improved texture networks:
  Maximizing quality and diversity in feed-forward stylization and texture
  synthesis,'' in \emph{Proceedings of the IEEE Conference on Computer Vision
  and Pattern Recognition}, 2017, pp. 6924--6932.

\bibitem{he2016identity}
K.~He, X.~Zhang, S.~Ren, and J.~Sun, ``Identity mappings in deep residual
  networks,'' in \emph{European conference on computer vision}.\hskip 1em plus
  0.5em minus 0.4em\relax Springer, 2016, pp. 630--645.

\bibitem{sankaranarayanan2015video}
A.~C. Sankaranarayanan, L.~Xu, C.~Studer, Y.~Li, K.~F. Kelly, and R.~G.
  Baraniuk, ``Video compressive sensing for spatial multiplexing cameras using
  motion-flow models,'' \emph{SIAM Journal on Imaging Sciences}, vol.~8, no.~3,
  pp. 1489--1518, 2015.

\bibitem{Set12}
B.~M. Roth~S., ``Fields of experts,'' in \emph{IEEE Conference on Computer
  Vision and Pattern Recognition}, 2009, p. 205.

\bibitem{yeh2015experimental}
L.~Yeh, J.~Dong, J.~Zhong, L.~Tian, M.~Chen, G.~Tang, M.~Soltanolkotabi, and
  L.~Waller, ``Experimental robustness of fourier ptychography phase retrieval
  algorithms,'' \emph{Optics Express}, vol.~23, no.~26, pp. 33\,214--33\,240,
  2015.

\bibitem{candes2015phase}
E.~J. Candes, X.~Li, and M.~Soltanolkotabi, ``Phase retrieval via wirtinger
  flow: Theory and algorithms,'' \emph{IEEE Transactions on Information
  Theory}, vol.~61, no.~4, pp. 1985--2007, 2015.

\bibitem{martin2001a}
D.~Martin, C.~Fowlkes, D.~Tal, and J.~Malik, ``A database of human segmented
  natural images and its application to evaluating segmentation algorithms and
  measuring ecological statistics,'' in \emph{Proceedings Eighth IEEE
  International Conference on Computer Vision. ICCV 2001}, vol.~2.\hskip 1em
  plus 0.5em minus 0.4em\relax IEEE, 2001, pp. 416--423.

\bibitem{li2013efficient}
C.~Li, W.~Yin, H.~Jiang, and Y.~Zhang, ``An efficient augmented lagrangian
  method with applications to total variation minimization,''
  \emph{Computational Optimization and Applications}, vol.~56, no.~3, pp.
  507--530, 2013.

\bibitem{zhang2020optimization}
J.~Zhang, C.~Zhao, and W.~Gao, ``Optimization-inspired compact deep compressive
  sensing,'' \emph{IEEE Journal of Selected Topics in Signal Processing},
  vol.~14, no.~4, pp. 765--774, 2020.

\bibitem{huang2011efficient}
J.~Huang, S.~Zhang, and D.~Metaxas, ``Efficient mr image reconstruction for
  compressed mr imaging,'' \emph{Medical Image Analysis}, vol.~15, no.~5, pp.
  670--679, 2011.

\bibitem{eksioglu2016decoupled}
E.~M. Eksioglu, ``Decoupled algorithm for mri reconstruction using nonlocal
  block matching model: Bm3d-mri,'' \emph{Journal of Mathematical Imaging and
  Vision}, vol.~56, no.~3, pp. 430--440, 2016.

\bibitem{pauluzzi2000comparison}
D.~R. Pauluzzi and N.~C. Beaulieu, ``A comparison of snr estimation techniques
  for the awgn channel,'' \emph{IEEE Transactions on communications}, vol.~48,
  no.~10, pp. 1681--1691, 2000.

\bibitem{riffe2007snr}
M.~Riffe, M.~Blaimer, K.~Barkauskas, J.~Duerk, and M.~Griswold, ``Snr
  estimation in fast dynamic imaging using bootstrapped statistics,'' in
  \emph{Proc Intl Soc Mag Reson Med}, vol. 1879, 2007.

\bibitem{dietrich2007measurement}
O.~Dietrich, J.~G. Raya, S.~B. Reeder, M.~F. Reiser, and S.~O. Schoenberg,
  ``Measurement of signal-to-noise ratios in mr images: influence of
  multichannel coils, parallel imaging, and reconstruction filters,''
  \emph{Journal of Magnetic Resonance Imaging: An Official Journal of the
  International Society for Magnetic Resonance in Medicine}, vol.~26, no.~2,
  pp. 375--385, 2007.

\end{thebibliography}

\end{document}